\documentclass[12pt,english]{elsarticle}
\usepackage[T1]{fontenc}
\usepackage[a4paper]{geometry}
\usepackage{color}
\usepackage{array}
\usepackage{float}
\usepackage{enumitem}
\usepackage{booktabs}
\usepackage{textcomp}
\usepackage{multirow}
\usepackage{amsmath}
\usepackage{amssymb}
\usepackage{graphicx}
\usepackage{stmaryrd}
\usepackage{mfirstuc}
\usepackage{setspace}
\usepackage{bbm}
\usepackage[colorlinks,allcolors=blue]{hyperref}
\usepackage{cleveref}
\usepackage{algorithm}
\usepackage{algpseudocode}
\usepackage{threeparttable}
\journal{Ocean Engineering}

\newcommand{\indicator}[1]{\mathbbm{1}_{#1}}

\newtheorem{assumption}{Assumption}
\newtheorem{remark}{Remark}

\begin{document}
	\setstretch{1.25}
	\begin{frontmatter}
		
		\title{Time-Frequency Mode Decomposition for Wind Turbine Vibration Monitoring under Variable Speed Operation}

		\author[add1,add2]{Wei~Zhou}
		\ead{zhouw6@szu.edu.cn}
		
		\author[add1,add2]{Wei-Jian~Li}
		\ead{weijianli@szu.edu.cn}
		
		\author[add1,add2]{Desen~Zhu}
		\ead{desenzhu@foxmail.com}
		
		\author[add3]{Hongbin~Xu}
		\ead{xuhongbin@semi.ac.cn}
		
		\author[add1,add2]{Wei-Xin~Ren\corref{cor1}}
		\ead{renwx@szu.edu.cn}
		
		\address[add1]{College of Civil and Transportation Engineering, Shenzhen University, Shenzhen, 518061, China}
		\address[add2]{National Key Laboratory of Green and Long-Life Road Engineering in Extreme Environment (Shenzhen), Shenzhen University, 518060, China}
		\address[add3]{Optoelectronic System Laboratory, Institute of Semiconductors, Chinese Academy of Sciences, Beijing 100083, China}
		
		\cortext[cor1]{Corresponding author.}
		
		\begin{abstract}
			Wind turbine vibration monitoring under variable speed operation requires separating nonstationary rotor-order components whose frequencies and operating intervals depend on operating state. These components can occupy local support regions in the short-time Fourier transform (STFT) plane rather than fixed spectral bands or continuous ridges. This study presents time-frequency mode decomposition (TFMD), a segmentation-based method that estimates connected STFT support regions and reconstructs one mode from each region. TFMD selects STFT coefficients with high magnitude, groups them by connected component labeling, filters small regions, expands retained support regions with mask dilation and conflict resolution, and reconstructs modes by inverse STFT. In a synthetic response with six operating states, TFMD separates the components of each state and produces low reconstruction error without specifying the number of components in advance. In a controlled wind turbine blade strain experiment, the first decomposition reconstructs nine modes whose peak frequencies lie near the nominal once per revolution frequencies and whose energies are concentrated in the corresponding operating intervals. Residual decomposition further reveals weaker harmonic structure. These results support TFMD as a practical candidate for vibration analysis under variable speed operation, while offshore field use requires validation under environmental loading and with measured operating references.
		\end{abstract}
		
	\end{frontmatter}
	
	\doublespacing
	
	\section{Introduction}
	\label{sec:introduction}
	
	Variable speed operation causes wind turbine vibration spectra to vary with rotor speed and operating state. Rotor-order components are therefore not fixed spectral lines. In stepped speed tests, a nominal order can appear in the short-time Fourier transform (STFT) plane as local support regions during the corresponding operating intervals, while weaker harmonics may be masked by stronger responses. Offshore wind turbine responses are further shaped by aerodynamic, hydrodynamic, drivetrain, and support structure interactions \citep{Jahani2022OWTDynamics}. A field study shows that environmental and harmonic excitations contribute differently to measured vibration and that rotor rotational frequency is relevant to structural safety assessment \citep{Dong2019VibrationSource}. These observations motivate decomposition methods that separate local components tied to operating states, tolerate noise, and retain weak harmonic components.

	More broadly, component separation supports system identification \citep{zhou2020modal,shang2024time,LIU2025108234}, feature extraction \citep{wang2025damage}, and fault diagnosis \citep{chauhan2024adaptive,wu2024quaternion,sun2025multivariate} in structural health monitoring and rotating machinery diagnostics.

	Offshore monitoring also faces practical constraints because environmental loading, operating variability, limited access, and maintenance cost shape condition-based maintenance strategies \citep{Scheu2019FMECA}. Field measurements over long periods have been used for operational modal analysis and structural vibration monitoring \citep{Devriendt2014AutomatedOMA,Dong2018StructuralVibration}, and monitoring performance can depend strongly on environmental and operational conditions \citep{Xiang2024OWTHealthMonitoring}. In wind turbine drivetrains, shaft speed information is often embedded in vibration responses through the fundamental rotational frequency and its harmonics, motivating order tracking and tacholess order tracking when direct speed measurements are unavailable \citep{He2016OrderTracking,Hou2019TacholessOrderTracking}. Decomposition outputs should therefore be interpretable against operating states and expected rotor-order frequencies, not only as stationary spectra.

	Existing methods differ in how they treat the number of modes, spectral support, and instantaneous frequency structure. Empirical mode decomposition (EMD) \citep{huang1998empirical} and its variants \citep{wu2009ensemble,torres2011complete} are adaptive but can suffer from mode mixing, end effects, and limited mathematical structure. Variational mode decomposition (VMD) \citep{dragomiretskiy2013variational} provides a more structured model, but it typically requires the number of modes in advance and assumes that modes are organized around central frequencies, which can be limiting for rapidly varying instantaneous frequencies. Variational nonlinear chirp mode decomposition relaxes this assumption by estimating time-varying instantaneous frequencies, but depends strongly on the initial instantaneous frequency estimates \citep{chen2017nonlinear}.

	Fourier spectral decomposition methods define bandpass filters in the global frequency domain, including the empirical wavelet transform \citep{Gilles2013Empirical}, Fourier decomposition method \citep{Singh2017Fourier}, empirical Fourier decomposition \citep{Zhou2022Empirical}, and Ramanujan Fourier mode decomposition \citep{Cheng2021Ramanujan}. They can be effective for stationary and mildly nonstationary signals, but global spectral segmentation is less suited to components whose supports are local, time-varying, or partially overlapping across operating intervals.

	Time-frequency reassignment and methods based on ridges instead sharpen or extract ridges from STFT or wavelet coefficients. Synchrosqueezing transform (SST) \citep{Daubechies2011Synchrosqueezed}, second-order SST \citep{oberlin2015second}, multisynchrosqueezing transform \citep{yu2018multisynchrosqueezing}, and synchroextracting transform (SET) and its variants \citep{Yu2017Synchroextracting,yu2021second,Ma2024Synchro} improve concentration or extract coefficients associated with instantaneous frequency ridges. Related methods, including variational generalized nonlinear mode decomposition (VGNMD) and its improved versions \citep{wang2024variational,Wang2025Improved} and nonlinear chirp mode extraction \citep{Xu2025Nonlinear}, use ridge information or demodulation to guide more specialized decomposition. These approaches are effective for many nonstationary signals, but their requirements for separated ridges, parameter tuning, or higher computation become restrictive under strong noise or ridge crossings, or when rotor-order components appear only during selected operating states.

	Across these families, many methods require a prescribed number of components, fixed or global frequency bands, parametric instantaneous frequency functions, or continuous time-frequency ridges. These assumptions are restrictive when diagnostically relevant vibration components form local support regions tied to operating states in the sampled STFT plane. This motivates reconstruction from connected support regions rather than from prescribed instantaneous frequency curves or manually selected ridges.

	To address this gap, this study introduces time-frequency mode decomposition (TFMD), a segmentation-based method for nonstationary vibration responses whose components occupy connected support regions in the sampled STFT plane. TFMD treats each mode as a time domain signal reconstructed from one connected support region, which may represent a persistent harmonic, a local rotor-order component, or another separated energy concentration. The procedure selects STFT coefficients with high magnitude, groups adjacent selected coefficients into connected regions, filters small regions, expands the retained support regions with mask dilation and conflict resolution, and synthesizes one mode from each final mask by inverse STFT.
	
	The paper proceeds as follows. \Cref{sec:methodology} establishes the TFMD formulation and segmentation procedure. \Cref{sec:Numerical_investigation} evaluates synthetic multicomponent signals through reconstruction, noise robustness, parameter sensitivity, and benchmark comparisons. \Cref{sec:application} examines a variable speed wind turbine blade strain response. \Cref{sec:discussion} discusses practical implications and limitations, and \Cref{sec:conclusion} summarizes the findings.
	
\section{Methodology}
\label{sec:methodology}

\begin{figure}[htbp]
    \centering
    \includegraphics[width=1\textwidth]{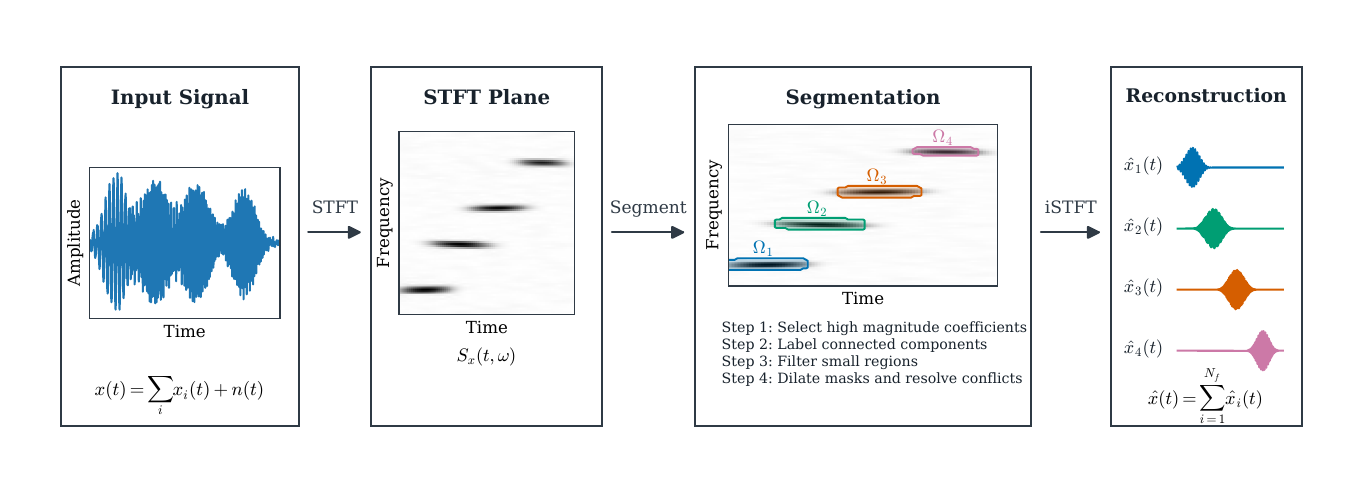}
    \caption{Conceptual overview of the TFMD segmentation and reconstruction workflow.}
    \label{fig:tfmd_segmentation_schematic}
\end{figure}

TFMD decomposes a signal by estimating support regions in the sampled STFT plane and reconstructing one mode from the coefficients retained in each region. The STFT and inverse STFT define the analysis and synthesis pair, while segmentation defines the regions used for reconstruction. In the proposed implementation, segmentation consists of four operations on the sampled STFT plane, namely coefficient selection by magnitude, connected component labeling, filtering by region size, and mask dilation with conflict resolution. A conceptual overview is shown in \Cref{fig:tfmd_segmentation_schematic}.

\subsection{Mode Reconstruction from Coefficient Support Regions}
\label{subsec:general_framework}

Let $x(t)\in L^2(\mathbb{R})$ be a multicomponent signal expressed as
\begin{equation}
    x(t)=\sum_{i=1}^{N}x_i(t)+n(t),
    \label{eq:signal_model}
\end{equation}
where $x_i(t)$ is the $i$-th component and $n(t)$ denotes additive noise. Let $\mathcal{T}:L^2(\mathbb{R})\rightarrow L^2(\mathbb{P})$ be a linear analysis operator that maps $x(t)$ to a coefficient field
\begin{equation}
    S_x(\mathbf{p})=(\mathcal{T}x)(\mathbf{p}), \qquad \mathbf{p}\in\mathbb{P},
\end{equation}
where $\mathbb{P}$ is the analysis domain. The corresponding synthesis operator $\mathcal{R}$ is assumed to provide perfect reconstruction for the unmasked coefficient field, i.e., $\mathcal{R}\mathcal{T}x=x$ under the selected analysis and synthesis settings.

TFMD seeks a collection of support regions $\{\Omega_i\}_{i=1}^{N_f}$ in $\mathbb{P}$, where $N_f$ is the number of modes identified in one decomposition stage. The binary mask associated with $\Omega_i$ is
\begin{equation}
    \mathcal{M}_i(\mathbf{p})=\mathbbm{1}_{\Omega_i}(\mathbf{p}).
\end{equation}
The $i$-th reconstructed mode is obtained by synthesizing only the coefficients selected by this mask:
\begin{equation}
    \hat{x}_i(t)=\mathcal{R}\left(S_x(\mathbf{p})\mathcal{M}_i(\mathbf{p})\right).
    \label{eq:general_recon}
\end{equation}
The decomposition problem is thus posed as estimating support regions in the analysis domain.

This segmentation formulation is meaningful when the coefficient field produced by the selected analysis operator satisfies two conditions. First, different components should be sufficiently separated in the analysis domain.
\begin{assumption}[Limited Overlap in the Analysis Domain]
    \label{assum:separability}
    For distinct components $x_i$ and $x_j$, their normalized coefficient overlap satisfies
    \begin{equation}
        \rho_{ij}=
        \frac{\int_{\mathbb{P}} |(\mathcal{T}x_i)(\mathbf{p})|\, |(\mathcal{T}x_j)(\mathbf{p})|\, d\mathbf{p}}
        {\left(\int_{\mathbb{P}} |(\mathcal{T}x_i)(\mathbf{p})|^2 d\mathbf{p}\right)^{1/2}
        \left(\int_{\mathbb{P}} |(\mathcal{T}x_j)(\mathbf{p})|^2 d\mathbf{p}\right)^{1/2}}
        \leq \varepsilon_{\mathrm{sep}},
        \label{eq:separability}
    \end{equation}
    where $0<\varepsilon_{\mathrm{sep}}\ll 1$.
\end{assumption}
Second, each component should have sufficient contrast relative to the background noise in its essential support.
\begin{assumption}[Energy Concentration]
    \label{assum:energy}
    Let $\Omega_i^{\mathrm{ess}}\subset\mathbb{P}$ denote the essential support of component $x_i$. The average component energy density in this support is assumed to dominate the expected noise energy density:
    \begin{equation}
        \frac{1}{|\Omega_i^{\mathrm{ess}}|}
        \int_{\Omega_i^{\mathrm{ess}}}|(\mathcal{T}x_i)(\mathbf{p})|^2d\mathbf{p}
        \gg
        E\left[|(\mathcal{T}n)(\mathbf{p})|^2\right],
        \label{eq:energy-concentration}
    \end{equation}
    where $E[\cdot]$ denotes expectation over the noise ensemble.
\end{assumption}
Assumption~\ref{assum:separability} limits component overlap, while Assumption~\ref{assum:energy} requires sufficient contrast between selected initial regions and the noise floor. If these conditions are violated, TFMD may merge nearby components or leave weak components in the residual.

\subsection{STFT Analysis and Synthesis}
\label{subsec:stft_analysis}

In this work, $\mathcal{T}$ and $\mathcal{R}$ are implemented by the STFT and inverse STFT. Let $g(t)\in L^2(\mathbb{R})$ be a real-valued analysis window with $\|g\|_2=1$. The continuous STFT is written as
\begin{equation}
    S_x(t,\omega)=(\mathcal{T}x)(t,\omega)
    =\int_{-\infty}^{\infty}x(\tau)g(\tau-t)e^{-j\omega\tau}d\tau,
    \label{eq:stft_def}
\end{equation}
with analysis domain $\mathbb{P}=\mathbb{R}^2$. Under the selected STFT convention, a corresponding continuous synthesis operation can be represented as
\begin{equation}
    (\mathcal{R}S)(t)=\frac{1}{2\pi}\iint_{\mathbb{R}^2}S(\tau,\omega)g(t-\tau)e^{j\omega t}d\omega d\tau.
    \label{eq:istft_def}
\end{equation}
In the discrete implementation, inverse STFT is performed using the selected synthesis window and overlap-add normalization, so the unmasked coefficient field reconstructs the input signal. By linearity of the STFT,
\begin{equation}
    S_x(t,\omega)=\sum_{i=1}^{N}S_{x_i}(t,\omega)+S_n(t,\omega).
\end{equation}
TFMD therefore estimates component support regions in the STFT plane and reconstructs modes from the selected coefficients by inverse STFT.

\subsection{Time-Frequency Segmentation}
\label{subsec:mask_construction}

The discrete STFT is computed with FFT size $K$, hop size $H$, sampling frequency $f_s$, and $M$ time frames. The time and frequency increments are $\Delta t=H/f_s$ and $\Delta f=f_s/K$, respectively. For real-valued signals, segmentation is performed on the nonnegative frequency half spectrum,
\begin{equation}
    \Lambda_+=\{(m,k):0\leq m<M,\;0\leq k\leq K_+\},
    \qquad K_+=\lfloor K/2\rfloor,
\end{equation}
where $(m,k)$ indexes the sampled time-frequency bin $(m\Delta t,k\Delta f)$. The one-sided STFT coefficients are denoted by $S_+[m,k]$. For any finite set $A\subseteq\Lambda_+$, $|A|$ denotes the number of STFT bins in $A$.

\subsubsection{Initial Coefficient Selection}
\label{subsubsec:core_init}

The first step identifies STFT bins whose magnitude is more consistent with component energy than with background noise. Let
\begin{equation}
    A[m,k]=\frac{|S_+[m,k]|}{\max_{(u,v)\in\Lambda_+}|S_+[u,v]|+\epsilon}
\end{equation}
be the normalized magnitude, with a small $\epsilon>0$ used only to avoid division by zero. TFMD applies two-cluster $k$-means clustering to the scalar samples $A[m,k]$:
\begin{equation}
    \min_{\{\ell_{m,k}\},\{\mu_c\}_{c=1}^{2}}
    \sum_{(m,k)\in\Lambda_+}\left(A[m,k]-\mu_{\ell_{m,k}}\right)^2,
    \qquad \ell_{m,k}\in\{1,2\}.
\end{equation}
The cluster with the larger centroid is interpreted as the class dominated by signal energy. The initial binary set is therefore
\begin{equation}
    \mathcal{B}=\{(m,k)\in\Lambda_+:\ell_{m,k}=c_s\},
    \qquad c_s=\arg\max_{c\in\{1,2\}}\mu_c .
\end{equation}
This step provides an adaptive selection of STFT bins with high magnitude without prescribing the number of modes. 

\subsubsection{Connected Component Labeling and Filtering}
\label{subsubsec:region_filtering}

Connected component labeling with 8-connectivity is applied to $\mathcal{B}$, producing connected sets $\{C_p\}_{p=1}^{P}$. Small isolated regions are removed using a normalized area threshold:
\begin{equation}
    \Omega_i^{(0)}\in
    \left\{C_p:\frac{|C_p|}{|\Lambda_+|}\geq \sigma\right\},
    \label{eq:initial_regions}
\end{equation}
where $\sigma$ is the minimum area ratio, so $|C_p|/|\Lambda_+|$ measures the relative area of a connected component on the one-sided STFT lattice. The retained regions $\{\Omega_i^{(0)}\}_{i=1}^{N_f}$ define the initial support regions, and their count gives the number of identified modes $N_f$ at that stage. Thus, the number of identified modes is inferred from connected support regions rather than supplied as a direct input.

\subsubsection{Mask Dilation with Conflict Resolution}
\label{subsubsec:icd}

The initial support regions contain the strongest coefficients but may omit peripheral energy introduced by the STFT window. TFMD expands each region by morphological dilation while keeping the final masks mutually exclusive. For the $i$-th initial region, the maximum dilation radius is
\begin{equation}
    r_i=\max\left(1,\left\lfloor \beta\sqrt{\frac{|\Omega_i^{(0)}|}{\pi}}\right\rfloor\right),
    \label{eq:dilation_radius}
\end{equation}
where $\beta$ controls the dilation extent and the term containing the square root is the equivalent radius of a disk with the same number of STFT bins as $\Omega_i^{(0)}$.

Let $\Omega_i^{(l)}$ denote the support of mode $i$ after dilation iteration $l$, initialized by $\Omega_i^{(0)}$. In this study, the structuring element $D$ is a $3\times3$ square on the discrete STFT lattice. During dilation, $\mathcal{S}^{(l)}$ records bins contested by multiple regions and left unassigned, with $\mathcal{S}^{(0)}=\emptyset$. At iteration $l$, the occupied set is
\begin{equation}
    \mathcal{O}^{(l-1)}=\bigcup_{j=1}^{N_f}\Omega_j^{(l-1)} .
\end{equation}
The candidate expansion set for mode $i$ is
\begin{equation}
    \mathcal{A}_i^{(l)}=
    \begin{cases}
    \left(\Omega_i^{(l-1)}\oplus D\right)
    \setminus\left(\mathcal{O}^{(l-1)}\cup\mathcal{S}^{(l-1)}\right), & l\leq r_i,\\
    \emptyset, & l>r_i,
    \end{cases}
    \label{eq:icd_candidates}
\end{equation}
where $\oplus$ denotes morphological dilation. Bins requested by more than one mode in the same iteration are marked as conflicts,
\begin{equation}
    \mathcal{C}^{(l)}=\bigcup_{i\neq j}\left(\mathcal{A}_i^{(l)}\cap\mathcal{A}_j^{(l)}\right),
    \qquad
    \mathcal{S}^{(l)}=\mathcal{S}^{(l-1)}\cup\mathcal{C}^{(l)} .
    \label{eq:icd_conflict}
\end{equation}
Each mode then accepts only uncontested bins:
\begin{equation}
    \Omega_i^{(l)}=\Omega_i^{(l-1)}\cup\left(\mathcal{A}_i^{(l)}\setminus\mathcal{C}^{(l)}\right).
    \label{eq:icd_update}
\end{equation}
After $l=\max_i r_i$, the final one-sided support is $\Omega_i^{(\mathrm{final})}=\Omega_i^{(\max_i r_i)}$. This rule prevents overlap by leaving contested bins unassigned rather than forcing them into one of the neighboring masks.

\subsection{Mode Reconstruction and Residual}
\label{subsec:mode_reconstruction}

The final one-sided support is converted to a one-sided binary mask,
\begin{equation}
    \mathcal{M}_i^+[m,k]=\mathbbm{1}_{\Omega_i^{(\mathrm{final})}}(m,k),
    \qquad (m,k)\in\Lambda_+ .
\end{equation}
For real-valued reconstruction, Hermitian symmetry is imposed to obtain a mask over the full frequency grid:
\begin{equation}
    \mathcal{M}_i[m,k]=
    \begin{cases}
    \mathcal{M}_i^+[m,k], & 0\leq k\leq K_+,\\
    \mathcal{M}_i^+[m,K-k], & K_+<k<K.
    \end{cases}
    \label{eq:hermitian_mask}
\end{equation}
Let $S[m,k]$ denote the full-grid STFT coefficients obtained before retaining the one-sided spectrum. The masked STFT coefficients for mode $i$ are
\begin{equation}
    S_i[m,k]=S[m,k]\mathcal{M}_i[m,k],
    \label{eq:selected_coefficients}
\end{equation}
and the corresponding mode in the time domain is
\begin{equation}
    \hat{x}_i(t)=\mathrm{iSTFT}\left(S_i[m,k]\right).
    \label{eq:region_istft_recon}
\end{equation}
The TFMD reconstruction and residual for one stage are
\begin{equation}
    \hat{x}^{(1)}(t)=\sum_{i=1}^{N_f}\hat{x}_i(t),
    \qquad
    e^{(1)}(t)=x(t)-\hat{x}^{(1)}(t).
    \label{eq:total_reconstruction}
\end{equation}
The residual can include unassigned STFT coefficients, such as coefficients with low magnitude, contested bins, and weak components not separated in the first segmentation stage.

\begin{remark}[Residual Decomposition]
    For large energy disparities, TFMD can be reapplied to $e^{(1)}(t)$ to inspect weaker components after dominant modes are removed. This residual decomposition uses the same segmentation and reconstruction steps. It is not a separate model.
\end{remark}

\begin{algorithm}[htbp]
\caption{TFMD segmentation and reconstruction for one stage}
\label{alg:tfmd}
\begin{algorithmic}[1]
\Require Signal $x(t)$; STFT parameters $g$, $K$, $H$; segmentation parameters $\sigma$, $\beta$, and $D$.
\Ensure Reconstructed modes $\{\hat{x}_i(t)\}_{i=1}^{N_f}$ and residual $e^{(1)}(t)$.
\State Compute the STFT $S[m,k]$ and retain the one-sided coefficients $S_+[m,k]$.
\State Normalize magnitudes $A[m,k]$ and apply two-cluster $k$-means to obtain bins dominated by signal energy $\mathcal{B}$.
\State Apply 8-connected component labeling to $\mathcal{B}$ and remove components with area ratio smaller than $\sigma$.
\State Set the retained components as initial supports $\{\Omega_i^{(0)}\}_{i=1}^{N_f}$.
\State Compute each dilation radius $r_i$ using Eq.~\eqref{eq:dilation_radius} and initialize $\mathcal{S}^{(0)}=\emptyset$.
\For{$l=1,\ldots,\max_i r_i$}
    \State Form candidate expansion sets $\mathcal{A}_i^{(l)}$ using Eq.~\eqref{eq:icd_candidates}.
    \State Identify contested bins $\mathcal{C}^{(l)}$ and update $\mathcal{S}^{(l)}$ using Eq.~\eqref{eq:icd_conflict}.
    \State Update each support $\Omega_i^{(l)}$ using Eq.~\eqref{eq:icd_update}.
\EndFor
\State Extend each one-sided support by Hermitian symmetry to construct $\mathcal{M}_i[m,k]$.
\State Reconstruct $\hat{x}_i(t)=\mathrm{iSTFT}(S[m,k]\mathcal{M}_i[m,k])$ for $i=1,\ldots,N_f$.
\State Compute $\hat{x}^{(1)}(t)=\sum_i\hat{x}_i(t)$ and $e^{(1)}(t)=x(t)-\hat{x}^{(1)}(t)$.
\end{algorithmic}
\end{algorithm}

	\section{Numerical Investigation}
	\label{sec:Numerical_investigation}
	
	This section evaluates the proposed TFMD method using six synthetic signals with known ground truth through four experiments. Experiment 1 establishes baseline performance without noise. Experiment 2 assesses robustness to additive white Gaussian noise (AWGN) across input signal-to-noise ratios (SNRs) of 10--40 dB. Experiment 3 evaluates sensitivity to the dilation factor $\beta$ from 0.1 to 1.0. Experiment 4 compares TFMD with EMD \citep{huang1998empirical}, VMD \citep{dragomiretskiy2013variational}, SET \citep{Yu2017Synchroextracting}, adaptive chirp mode decomposition (ACMD) \citep{chen2019detection}, and VGNMD \citep{wang2024variational} in terms of reconstruction error, noise robustness, and computational efficiency. The synthetic set includes both generic nonstationary test signals and a synthetic signal with stepped operating states that links the numerical experiments to the laboratory wind turbine analysis in \Cref{sec:application}.
	
	\subsection{Datasets}
	\label{subsec:datasets_exp}
	
	\subsubsection{Synthetic Signals}
	\label{subsubsec:signal_synthesis}
	Six distinct synthetic signals, denoted $x_1(t)$ through $x_6(t)$, were designed to represent signal characteristics encountered in nonstationary vibration analysis. Cases 1--5 include frequency separated chirps, sinusoidal frequency modulation (FM), heterogeneous multicomponent mixtures, amplitude modulation (AM), transient components, and complex nonlinear components. Case 6 introduces a synthetic response with stepped operating states and six local components, each active in one successive 20 s state. Unless otherwise specified, signals were generated with a sampling frequency $f_s = 1000$ Hz and a duration of 1 s. Case 5 and Case 6 use the durations and sampling frequencies stated in their definitions.
	
	\paragraph{Signal Case 1: Frequency Separated Chirps}
	\label{subsubsec:case1_exp}
	Case 1 comprises two well separated chirp components, a linear chirp ($x_{1,1}$) and a quadratic chirp ($x_{1,2}$):
	\begin{equation} \label{eq:sig1_exp}
		x_1(t) = \underbrace{1.0 \cdot \cos(2\pi \cdot [20t + 25t^2])}_{x_{1,1}(t)} + \underbrace{0.9 \cdot \cos(2\pi \cdot [130t + \frac{50}{3}t^3])}_{x_{1,2}(t)}
	\end{equation}
	
	\paragraph{Signal Case 2: Sinusoidal FM Signals}
	\label{subsubsec:case2_exp}
	Case 2 consists of two sinusoidally frequency modulated components with distinct modulation characteristics:
	\begin{equation} \label{eq:sig2_exp}
		x_2(t) = \underbrace{1.2 \cdot \cos\left(2\pi \cdot 100t + 15\sin(2\pi \cdot 2t)\right)}_{x_{2,1}(t)} + \underbrace{1.0 \cdot \cos\left(2\pi \cdot 250t + 5\sin(2\pi \cdot 5t)\right)}_{x_{2,2}(t)}
	\end{equation}
	
	\paragraph{Signal Case 3: Four Component Mixture}
	\label{subsubsec:case3_exp}
	Case 3 combines heterogeneous signal types, including a linear chirp, a pure tone, a time limited FM signal, and a transient AM burst:
	\begin{align} \label{eq:sig3_exp}
		x_3(t) &= \underbrace{1.0 \cdot \cos(2\pi \cdot [10t + 15t^2])}_{x_{3,1}(t)} + \underbrace{0.9 \cdot \sin(2\pi \cdot 100t)}_{x_{3,2}(t)} \nonumber \\
		&+ \underbrace{1.1 \cdot \indicator{[0,0.7]}(t) \cdot \cos\left(2\pi \cdot 350t + 5\sin(2\pi \cdot 6t)\right)}_{x_{3,3}(t)} \nonumber \\
		&+ \underbrace{1.2 \cdot \indicator{[0.6,0.9]}(t) \cdot w(t) \cdot \sin(2\pi \cdot 200t)}_{x_{3,4}(t)}
	\end{align}
	where $w(t)$ represents a Tukey window with taper parameter 0.25.
	
	\paragraph{Signal Case 4: Low Frequency Chirp and AM Tone}
	\label{subsubsec:case4_exp}
	Case 4 assesses performance on signals combining FM and AM features:
	\begin{align} \label{eq:sig4_exp}
		x_4(t) &= \underbrace{1.0 \cdot \cos(2\pi \cdot [20t + 30t^2])}_{x_{4,1}(t)} + \underbrace{1.1 \cdot (0.8 + 0.4\cos(2\pi \cdot 2t)) \cdot \sin(2\pi \cdot 200t)}_{x_{4,2}(t)}
	\end{align}
	
	\paragraph{Signal Case 5: Generalized Nonlinear Signal}
	\label{subsubsec:case5_exp}
	Case 5 is a highly nonstationary signal with a duration of 3 s (3000 samples). It incorporates seven components, including nonlinear chirps and components defined by prescribed complex spectra $G_i(f)$:
	\begin{align} \label{eq:sig5_exp}
		x_5(t) &= \underbrace{\cos(2\pi \cdot (170t + 20t^2 + 3\cos(3\pi t)))}_{x_{5,1}(t)} + \underbrace{\indicator{[0,1.5]}(t) \cdot \cos(2\pi \cdot (75t + 20t^2))}_{x_{5,2}(t)} \nonumber \\
		&+ \underbrace{\indicator{[1,3]}(t) \cdot \cos(2\pi \cdot (10t + 20t^2 + 3\cos(3\pi t)))}_{x_{5,3}(t)} + \sum_{i=4}^{7} x_{5,i}(t)
	\end{align}
	The components $x_{5,i}(t)$ ($i=4 \dots 7$) are generated via the inverse Fourier transform of:
	\begin{align} \label{eq:case5_spectra}
		G_4(f) &= 30e^{-j2\pi(0.4f + 2\cos(2\pi f/100))} \cdot \indicator{[f_s/4, f_s/2)}(f) \nonumber \\
		G_5(f) &= 30e^{-j2\pi(0.8f + 0.0005f^2)} \cdot \indicator{[3f_s/10, f_s/2)}(f) \nonumber \\
		G_6(f) &= 30e^{-j2\pi(1.8f + 2\cos(2\pi f/100))} \cdot \indicator{[7f_s/20, f_s/2)}(f) \nonumber \\
		G_7(f) &= 30e^{-j2\pi(2.2f + 0.0005f^2)} \cdot \indicator{[4f_s/10, f_s/2)}(f)
	\end{align}
	
	\paragraph{Signal Case 6: Synthetic Signal with Stepped Operating States}
	\label{subsubsec:case6_exp}
	Case 6 is a synthetic signal with stepped operating states designed to test whether TFMD can recover separated connected support regions in the sampled STFT plane. It is not intended as a full turbine simulation. Instead, it provides a simplified controlled representation of variable speed operation in which one local component linked to the fundamental frequency is active in each operating state. The sampling frequency is 50 Hz and the duration is 120 s. The signal contains six operating states, each lasting 20 s, with piecewise constant fundamental frequencies
	\begin{equation} \label{eq:case6_freq}
		f_0(t) \in \{1.20,\,1.85,\,2.55,\,3.25,\,3.95,\,4.65\}\ \mathrm{Hz}.
	\end{equation}
	For state $k$, the local component is active only on $[20(k-1),20k)$ s:
	\begin{align} \label{eq:sig6_exp}
		x_6(t) &= \sum_{k=1}^{6} x_{6,k}(t), \\
		x_{6,k}(t) &= \indicator{[20(k-1),20k)}(t)\,a_k(t)\cos\theta(t), \\
		\theta(t) &= 2\pi\int_0^t f_0(\tau)\,d\tau .
	\end{align}
	The envelope is defined on the local interval by $a_k(t)=w_k(t)[1+0.04\cos(2\pi\cdot0.06(t-20(k-1))+0.35k)]$, where $w_k(t)$ is a unit peak Tukey window with taper parameter 0.30 supported on $[20(k-1),20k)$.
	
	\subsection{Performance Metrics}
	\label{subsec:metrics_exp}
	
	Quantitative assessment relies on the relative $L_2$ error and output SNR. For the $c$-th signal case $(c=1,...,6)$, let $x_{c,i}(t)$ and $\hat{x}_{c,i}(t)$ denote the ground truth component and the extracted mode for the $i$-th component, respectively. Reconstructed modes are paired with ground truth components using a greedy maximum absolute Pearson correlation rule: each ground truth component is assigned the still-unused reconstructed mode with the largest absolute correlation. The relative error for each component $\mathcal{E}_{\text{rel},i}$ and the total relative error $\mathcal{E}_{\text{rel, total}}$ are defined as:
	\begin{equation} \label{eq:errors_exp}
		\mathcal{E}_{\text{rel},i} = \frac{\| x_{c,i}(t) - \hat{x}_{c,i}(t) \|_2}{\| x_{c,i}(t) \|_2}, \quad
		\mathcal{E}_{\text{rel, total}} = \frac{\| x_c(t) - \hat{x}_c(t) \|_2}{\| x_c(t) \|_2}
	\end{equation}
	The average mode error is calculated after this assignment as $\mathcal{E}_{\text{rel, avg}} = \frac{1}{N} \sum_{i=1}^{N} \mathcal{E}_{\text{rel},i}$. Thus, $\mathcal{E}_{\text{rel, total}}$ and $\mathcal{E}_{\text{rel, avg}}$ measure full signal reconstruction and assigned mode accuracy, respectively.
	
	Performance under noise is quantified by the output SNR:
	\begin{equation} \label{eq:output_snr_metric_sym_exp}
		\text{SNR}_{\text{out}} = 10 \log_{10} \left( \frac{\| x_c(t) \|_2^2}{\| x_c(t) - \hat{x}_c(t) \|_2^2} \right) \quad \text{dB}
	\end{equation}

	\subsection{Implementation Details}
	\label{subsec:implementation_details_exp}
	
	For the proposed TFMD method, the STFT was computed using a discrete Gaussian window with a length of 128 samples and a shape parameter of 2.5. The transform used an FFT size of $K=256$ and an overlap ratio of 0.9. The $k$-means coefficient selection step used two clusters, a maximum of 100 iterations, one replicate, and no parallel execution within the $k$-means call. The segmentation process used a $3\times3$ square structuring element. The minimum area ratio parameter was set to $\sigma = 10^{-3}$, and the dilation factor was set to $\beta = 0.5$ for all cases unless otherwise specified in the sensitivity analysis.
	
	For the benchmark methods in Experiment 4, fixed parameter settings were used across the tested cases. For methods that accept or use an output count setting, the known number of components was supplied. VMD used this value as the number of modes, ACMD and SET extracted that many components or ridges, and EMD used it as the maximum number of intrinsic mode functions. VGNMD used its internal mode detection. TFMD does not use the number of components as a direct input. The number of modes is inferred from connected support regions identified under the selected STFT resolution and segmentation parameters. This protocol provides the benchmark methods with prior information unavailable in practice, thereby ensuring a conservative evaluation of TFMD's relative performance.
	
	All numerical simulations were executed using MATLAB R2024b on a workstation equipped with a 13th Gen Intel Core i7-13620H processor (2.40 GHz) and 40 GB of RAM. Runtime measurements in Experiment 4 used one untimed warm-up call per method before the timed sweep to exclude MATLAB first-call overhead. Each method-input pair was then timed five times, and the median runtime was used for that pair.

	\subsection{Results and Analysis}
	\label{subsec:results_analysis}

	The following subsections present the results of the four experiments outlined above.

	\subsubsection{Performance on Synthetic Signals without Noise}
	\label{subsec:results_exp1_clean}
	
	Baseline decomposition performance was first evaluated without added noise. Quantitative performance metrics are summarized in \Cref{tab:exp1_results_overall_transposed} and \Cref{tab:exp1_results_individual_augmented}. For all six cases, the algorithm identified the correct number of modes ($N_f$), matching the ground truth number of components ($N$) without using the number of components as a direct input.
	
	\begin{table}[htbp]
		\centering
		\caption{Overall reconstruction performance of TFMD (Experiment 1). The table reports the ground truth number of components ($N$) versus the identified number of modes ($N_f$) and the total relative error.}
		\label{tab:exp1_results_overall_transposed}
		\vspace{1ex}
		\small 
		\setlength{\tabcolsep}{4pt} 
		\begin{tabular}{lcccccc}
			\toprule
			Metric & Case 1 & Case 2 & Case 3 & Case 4 & Case 5 & Case 6 \\
			\midrule
			$N / N_f$ & 2 / 2 & 2 / 2 & 4 / 4 & 2 / 2 & 7 / 7 & 6 / 6 \\
			$\mathcal{E}_{\text{rel, total}}$ ($\times 10^{-2}$) & 2.09 & 3.01 & 3.38 & 4.44 & 3.62 & 0.12 \\
			\bottomrule
		\end{tabular}
	\end{table}
	
	\begin{table}[htbp]
		\centering
		\caption{TFMD individual mode decomposition performance for Experiment 1.}
		\label{tab:exp1_results_individual_augmented}
		\vspace{1ex}
		\small
		\setlength{\tabcolsep}{3.5pt} 
		\begin{tabular}{ccccccccc}
			
			\toprule
			Case  & $\mathcal{E}_{\text{rel, avg}}$  & $\mathcal{E}_{\text{rel},1}$ & $\mathcal{E}_{\text{rel},2}$ & $\mathcal{E}_{\text{rel},3}$ & $\mathcal{E}_{\text{rel},4}$ & $\mathcal{E}_{\text{rel},5}$ & $\mathcal{E}_{\text{rel},6}$ & $\mathcal{E}_{\text{rel},7}$ \\
			& ($\times 10^{-2}$) & ($\times 10^{-2}$) & ($\times 10^{-2}$) & ($\times 10^{-2}$) & ($\times 10^{-2}$) & ($\times 10^{-2}$) & ($\times 10^{-2}$) & ($\times 10^{-2}$) \\
			\midrule
			1 & 2.59 & 1.53 & 3.64 & ---  & ---  & ---  & ---  & ---  \\
			2 & 3.64 & 2.89 & 4.38 & ---  & ---  & ---  & ---  & ---  \\
			3 & 4.68 & 4.31 & 6.13 & 5.81 & 2.48 & ---  & ---  & ---  \\
			4 & 5.01 & 2.70 & 7.32 & ---  & ---  & ---  & ---  & ---  \\
			5 & 5.64 & 3.22 & 3.96 & 3.34 & 5.68 & 6.24 & 7.42 & 9.63 \\
			6 & 0.18 & 0.15 & 0.21 & 0.21 & 0.21 & 0.17 & 0.12 & --- \\
			\bottomrule
		\end{tabular}
	\end{table}
	
\begin{figure}[htbp]
		\centering
		\includegraphics[width=1\textwidth]{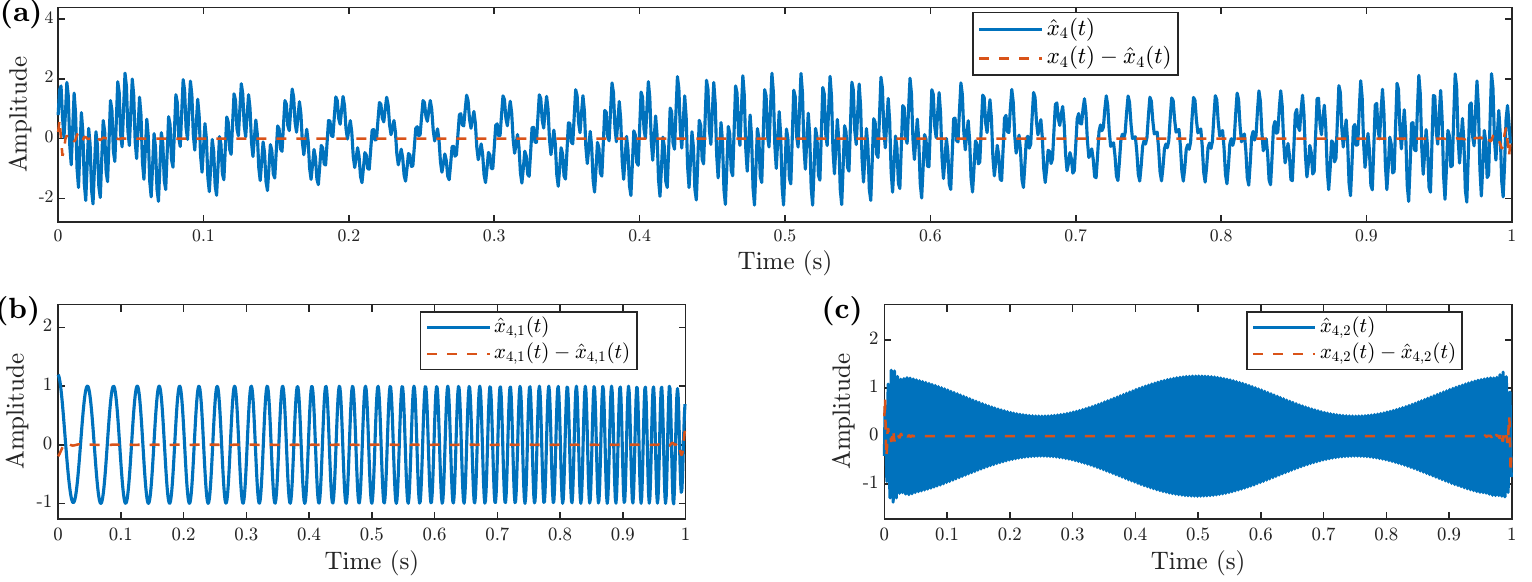}
		\caption{TFMD decomposition results for $x_4(t)$ without noise. (a) Total reconstructed signal $\hat{x}_4(t)$ and error. (b) First extracted mode $\hat{x}_{4,1}(t)$ and estimation error. (c) Second extracted mode $\hat{x}_{4,2}(t)$ and estimation error.}
		\label{fig:exp1_case4_results_fig}
	\end{figure}

	Detailed decomposition accuracy is provided in \Cref{tab:exp1_results_individual_augmented}. The average mode error does not exceed approximately $5.64\times10^{-2}$ across the six cases, with the lowest value obtained in Case 6, the synthetic signal with stepped operating states. In the representative Case 4 example shown in \Cref{fig:exp1_case4_results_fig}, the extracted modes $\hat{x}_{4,1}(t)$ and $\hat{x}_{4,2}(t)$ closely match the ground truth components, and the minor reconstruction deviations are primarily concentrated at the signal boundaries, a characteristic attributable to the windowing effects inherent in the STFT framework.

	\subsubsection{Noise Robustness Assessment}
	\label{subsec:results_exp2_noise}
	
\begin{figure}[htbp]
		\centering
		\includegraphics[width=0.9\textwidth]{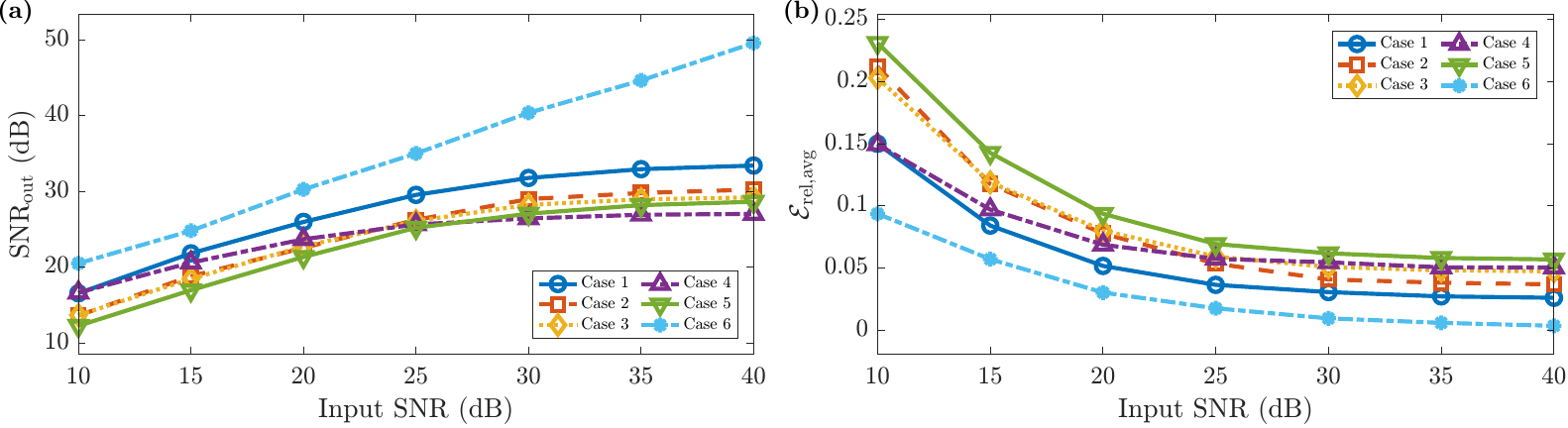} 
		\caption{TFMD noise robustness assessment. (a) $\mathrm{SNR}_{\mathrm{out}}$ versus input SNR. (b) Average mode error $\mathcal{E}_{\text{rel, avg}}$ versus input SNR.}
		\label{fig:exp2_tfmd_robustness_metrics}
	\end{figure}

	Experiment 2 evaluated robustness against AWGN across input SNRs of 10--40 dB. As illustrated in \Cref{fig:exp2_tfmd_robustness_metrics}(a), the output SNR increases with input SNR. At 10 dB input SNR, it remains above 12 dB for all six cases. Concurrently, the average mode error $\mathcal{E}_{\text{rel, avg}}$ shown in \Cref{fig:exp2_tfmd_robustness_metrics}(b) decreases as the input SNR increases from 10 to 40 dB for every case. TFMD identifies the correct number of modes for all six cases throughout the tested SNR range, matching the ground truth number of components under each noise condition.
	
	\subsubsection{Parameter Sensitivity Analysis}
	\label{subsec:results_exp3_sensitivity}
	
\begin{figure}[htbp]
		\centering
		\includegraphics[width=1\textwidth]{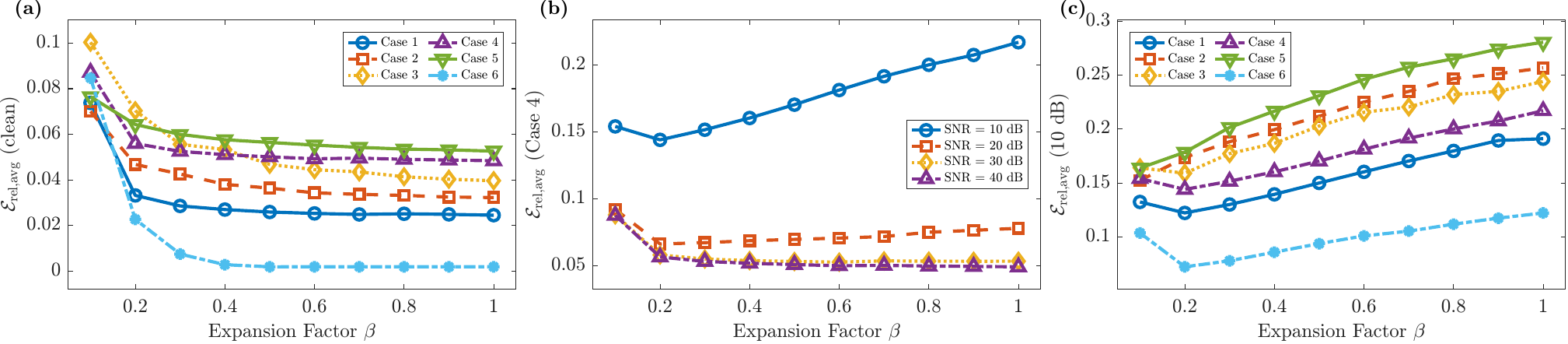}
		\caption{Parameter sensitivity analysis for dilation factor $\beta$. (a) $\mathcal{E}_{\text{rel, avg}}$ versus $\beta$ without noise for all six cases. (b) $\mathcal{E}_{\text{rel, avg}}$ versus $\beta$ for Case 4 across input SNR levels. (c) $\mathcal{E}_{\text{rel, avg}}$ versus $\beta$ at 10 dB for all six cases.}
		\label{fig:exp3_beta_sensitivity_fig}
	\end{figure}

	The sensitivity of TFMD to the dilation factor $\beta$ was evaluated by varying the parameter from 0.1 to 1.0. In the results without noise in \Cref{fig:exp3_beta_sensitivity_fig}(a), larger dilation factors reduce the average mode error for Cases 1--5, whereas $\beta=0.5$ is best for Case 6. For Cases 1--5, using $\beta=0.5$ increases the average mode error by at most 18.1\% relative to the best setting without noise. For Case 4, \Cref{fig:exp3_beta_sensitivity_fig}(b) shows that smaller dilation factors are favored at lower input SNRs. \Cref{fig:exp3_beta_sensitivity_fig}(c) extends this low SNR check to all six cases at 10 dB, where smaller dilation factors remain favored. The default value $\beta = 0.5$ is therefore used as a compromise across noise-free and low-SNR conditions rather than as a universally optimal value.
	
	\subsubsection{Comparative Analysis with Benchmark Methods}
	\label{subsec:results_exp4_comparison}
	
\begin{figure}[htbp]
		\centering
		\includegraphics[width=1\textwidth]{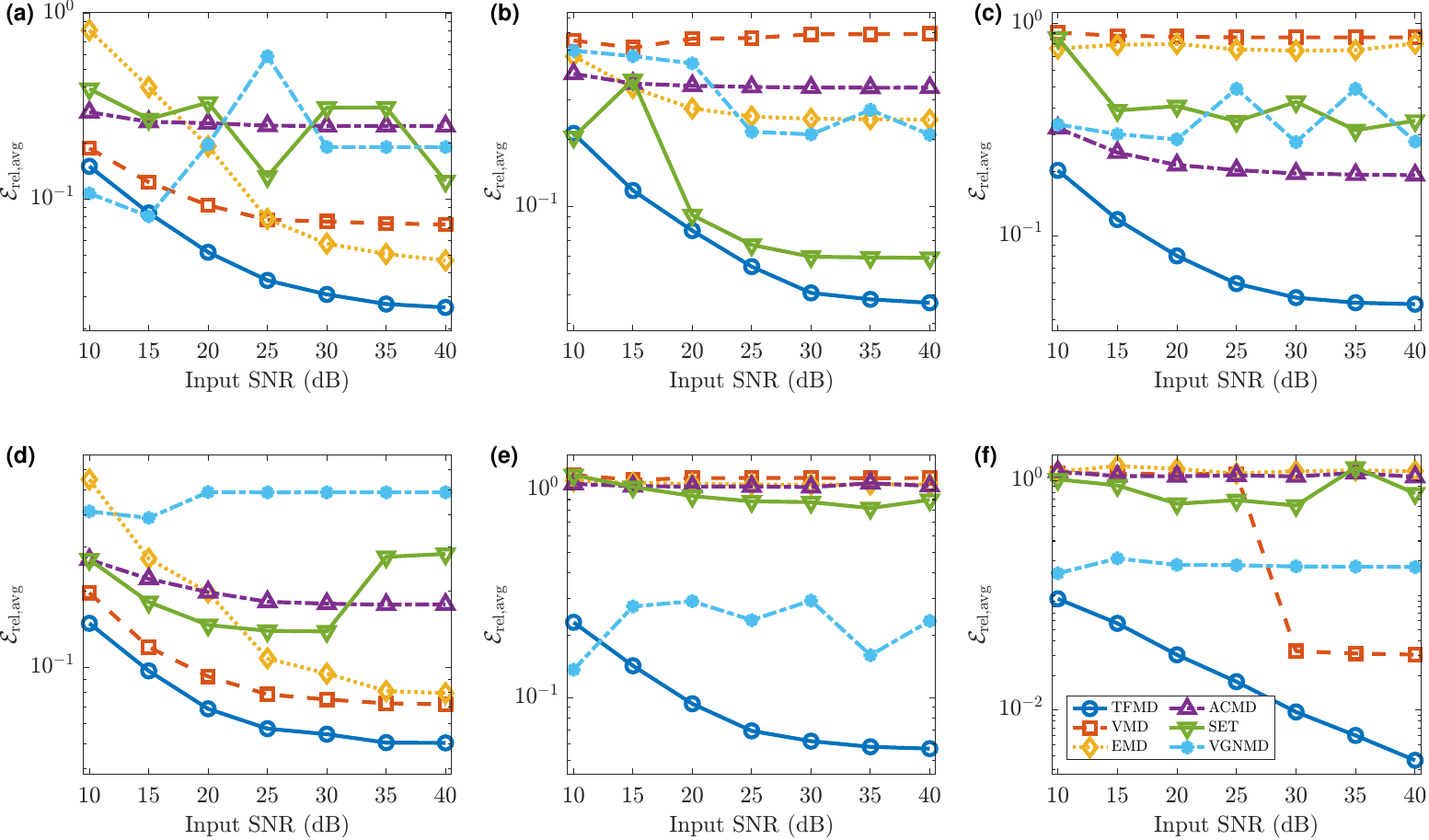}
		\caption{Comparative analysis of average mode error $\mathcal{E}_{\text{rel, avg}}$ as a function of input SNR. Each subplot corresponds to one of the six synthetic signal cases: (a) Case 1, (b) Case 2, (c) Case 3, (d) Case 4, (e) Case 5, (f) Case 6. The shared legend identifies the six methods.}
		\label{fig:exp4_avg_mode_error_comparison_fig}
	\end{figure}

	Experiment 4 benchmarks TFMD against EMD, VMD, ACMD, SET, and VGNMD, focusing on average mode error and computational efficiency. The evolution of the average mode error $\mathcal{E}_{\text{rel, avg}}$ as a function of input SNR is presented in \Cref{fig:exp4_avg_mode_error_comparison_fig}. Across the seven tested SNR levels, TFMD has the lowest median $\mathcal{E}_{\text{rel, avg}}$ in all six cases, but it is not the lowest method at every SNR in Cases 1, 2, and 5. After taking the median over the seven SNR levels, the closest benchmark $\mathcal{E}_{\text{rel, avg}}$ is 2.12, 1.25, 3.43, 1.36, 3.41, and 10.16 times the TFMD value in Cases 1--6, respectively.

	Among the benchmark methods, the closest benchmark varies with the signal case: VMD is closest by median $\mathcal{E}_{\text{rel, avg}}$ in Cases 1 and 4, SET in Case 2, ACMD in Case 3, and VGNMD in Cases 5 and 6. This variation indicates that no single benchmark method is consistently closest across the six synthetic cases.
	
	\begin{table}[htbp]
		\centering
		\caption{Average computational runtime (seconds) across all input SNR levels; each input runtime is the median of five repeated timed calls. \textbf{Bold} indicates the shortest runtime; underlining indicates the second shortest.}
		\label{tab:runtime_comparison}
		\small
		\setlength{\tabcolsep}{5pt}
		\begin{tabular}{lcccccc}
			\toprule
			Method & Case 1 & Case 2 & Case 3 & Case 4 & Case 5 & Case 6 \\
			\midrule
			TFMD   & 0.0148 & \underline{0.0095} & \underline{0.0158} & \underline{0.0089} & \underline{0.0365} & \underline{0.0573} \\
			VMD    & 0.0452 & 0.0426 & 0.0796 & 0.0459 & 0.2692 & 0.2696 \\
			EMD    & \textbf{0.0021} & \textbf{0.0005} & \textbf{0.0012} & \textbf{0.0006} & \textbf{0.0022} & \textbf{0.0030} \\
			ACMD   & \underline{0.0102} & 0.0113 & 0.0249 & 0.0110 & 0.7302 & 10.7765 \\
			SET    & 0.0666 & 0.0706 & 0.0930 & 0.0696 & 0.7958 & 3.1299 \\
			VGNMD  & 0.7097 & 1.1514 & 1.8651 & 0.7608 & 6.5738 & 13.6741 \\
			\bottomrule
		\end{tabular}
	\end{table}
	
	Computational efficiency, summarized in \Cref{tab:runtime_comparison}, further differentiates the practical utility of these methods. Under the median-of-five timing protocol, EMD gives the lowest runtimes in all six cases, while TFMD is second fastest in Cases 2--6. TFMD averages 0.0089--0.0573 s across the six cases. In Case 1, ACMD runs faster than TFMD. ACMD is also close to TFMD in Cases 2 and 4, but its runtime increases sharply in Cases 5 and 6. Thus, TFMD is not the uniformly fastest method. Its runtime remains moderate while retaining the advantages in median $\mathcal{E}_{\text{rel, avg}}$ reported in \Cref{fig:exp4_avg_mode_error_comparison_fig}.

	\section{Controlled Laboratory Evaluation of Variable Speed Wind Turbine Strain Response}
	\label{sec:application}
	
\begin{figure}[htbp]
		\centering
		\includegraphics[width=0.75\textwidth]{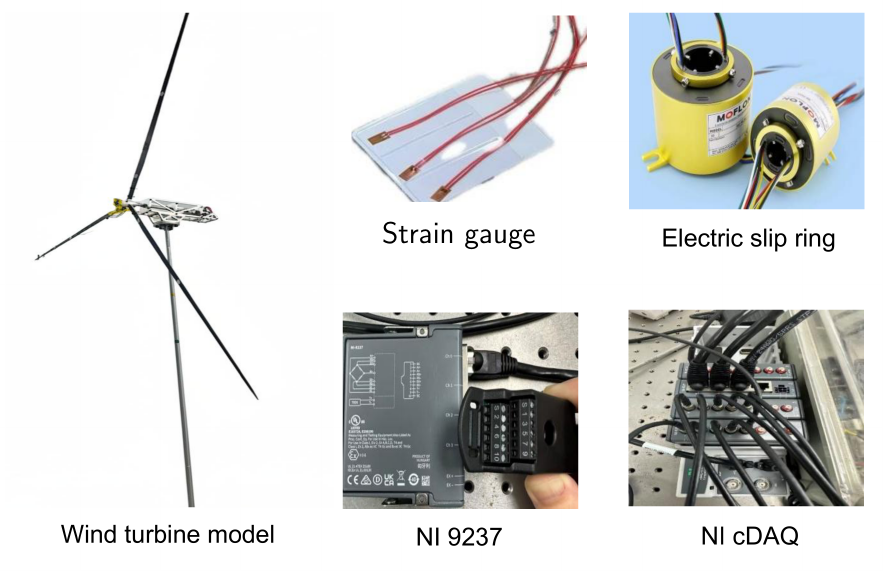}
		\caption{Key components of the experimental setup, including the laboratory scale wind turbine model, the strain gauge used for the analyzed channel, the electric slip ring for signal transmission, and the National Instruments CompactDAQ (cDAQ) chassis and NI 9237 module.}
		\label{fig:experimental_setup_components}
	\end{figure}

	To evaluate TFMD on a measured blade strain response associated with turbine vibration, a controlled laboratory experiment was conducted on a scaled horizontal axis wind turbine. The experiment is treated as laboratory evidence for vibration analysis under controlled operating states rather than as an offshore field demonstration. The setup, illustrated in \Cref{fig:experimental_setup_components}, features a rotor with three 0.751 m carbon fiber blades. The supporting structure consists of an aluminum alloy tower with a height of 1.095 m and a mass of 0.292 kg, topped by an aluminum alloy nacelle with a mass of 0.544 kg and dimensions of $0.214 \times 0.054 \times 0.054$ m. Flapwise bending strain was measured using a strain gauge positioned at one fifth of the blade length from the root. The analog signals were transmitted through a slip ring mounted on the rotor and acquired synchronously at a sampling rate of 2000 Hz using a National Instruments CompactDAQ system.
	
	To generate a controlled nonstationary strain response, the turbine was operated under a stepped speed protocol. The rotational speed was varied from 10 to 90 revolutions per minute (RPM) in steps of 10 RPM, with each speed maintained for 30 s. This protocol provides known nominal 1P (once per revolution) and 2P (twice per revolution) frequencies while retaining measurement imperfections, making it suitable for assessing whether extracted modes are localized in the corresponding operating intervals and whether weaker harmonics can be detected in the residual.
	
	\subsection{Signal Preprocessing}
	\label{subsec:signal_preprocessing}
	
\begin{figure}[htbp]
		\centering
		\includegraphics[width=0.95\textwidth]{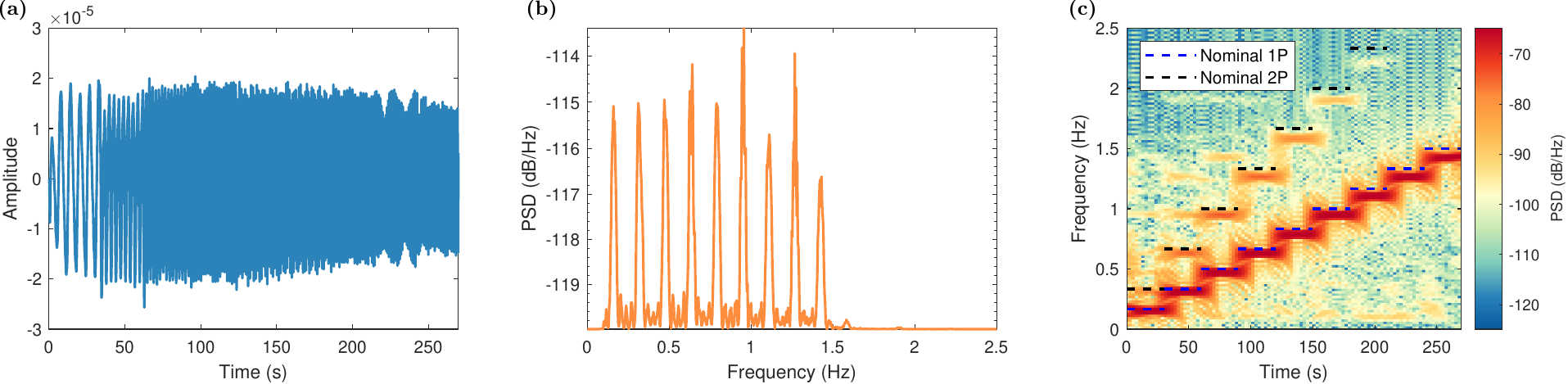}
		\caption{Analysis of the preprocessed strain signal. (a) Time domain waveform. (b) Power spectral density (PSD). (c) STFT spectrogram with overlaid nominal 1P and 2P frequencies.}
		\label{fig:input_signal_analysis}
	\end{figure}

	Prior to decomposition, the raw strain signal underwent a preprocessing pipeline to isolate the rotational response band containing the nominal 1P frequency range and the lower speed part of the 2P frequency range. The signal was first detrended and then filtered using a Butterworth bandpass filter of order 12 with cutoffs at 0.1 Hz and 1.5 Hz to eliminate drift below 0.1 Hz and response above 1.5 Hz. For the fixed-length STFT window used below, the filtered signal was downsampled to 5 Hz to refine the frequency-bin spacing while satisfying the Nyquist criterion. As shown in \Cref{fig:input_signal_analysis}, the resulting STFT spectrogram exhibits a distinct staircase pattern corresponding to the discrete operating states, with visible deviations between the observed spectral peaks and the nominal frequency lines at higher RPMs.
	
	\subsection{Decomposition Results}
	\label{subsec:application_results}
	
\begin{figure}[htbp]
		\centering
		\includegraphics[width=0.95\textwidth]{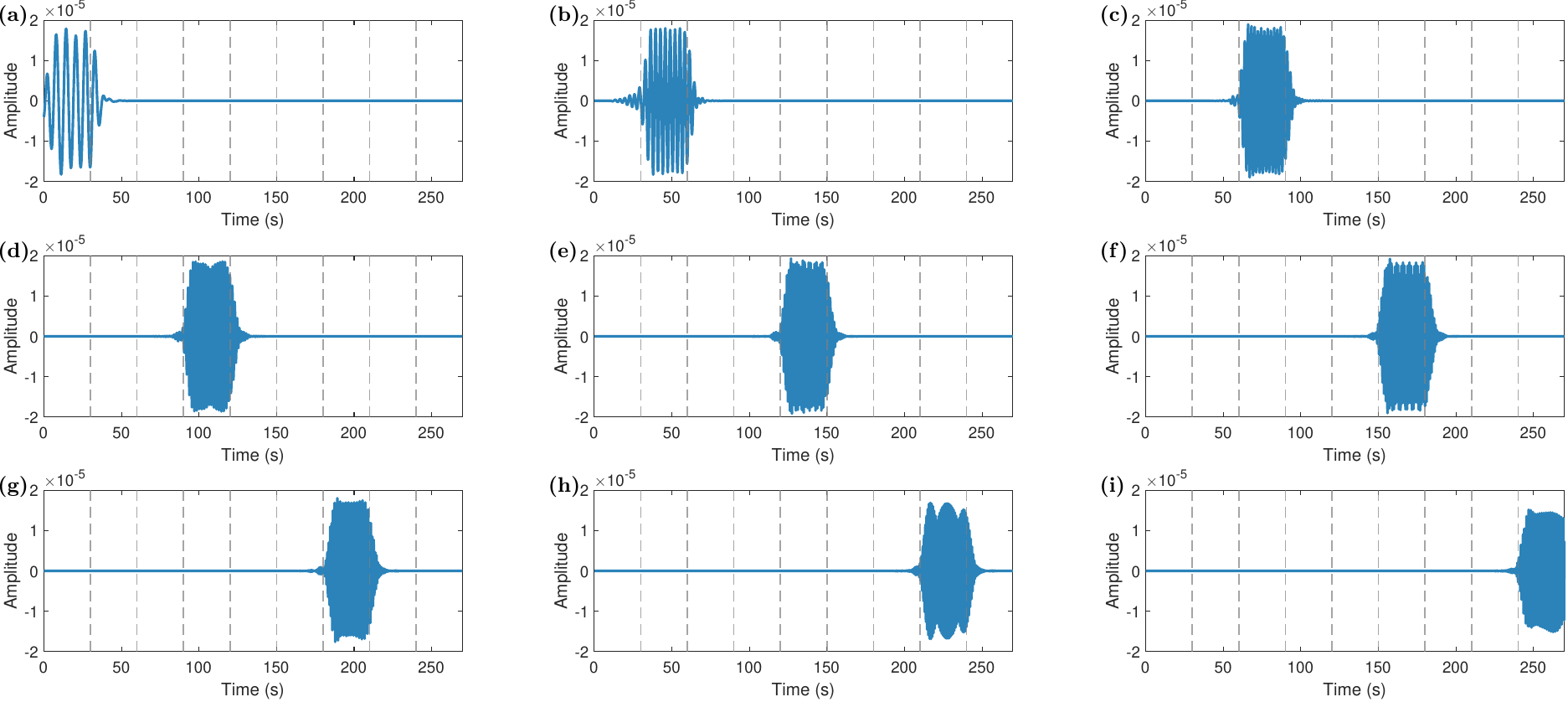}
		\caption{Time domain representations of Modes 1 through 9 extracted by TFMD. Subplots (a) through (i)
			correspond to Modes 1 through 9, respectively. Dashed vertical lines indicate 30 s operating intervals.}
		\label{fig:comp1_time}
	\end{figure}

\begin{figure}[htbp]
		\centering
		\includegraphics[width=0.95\textwidth]{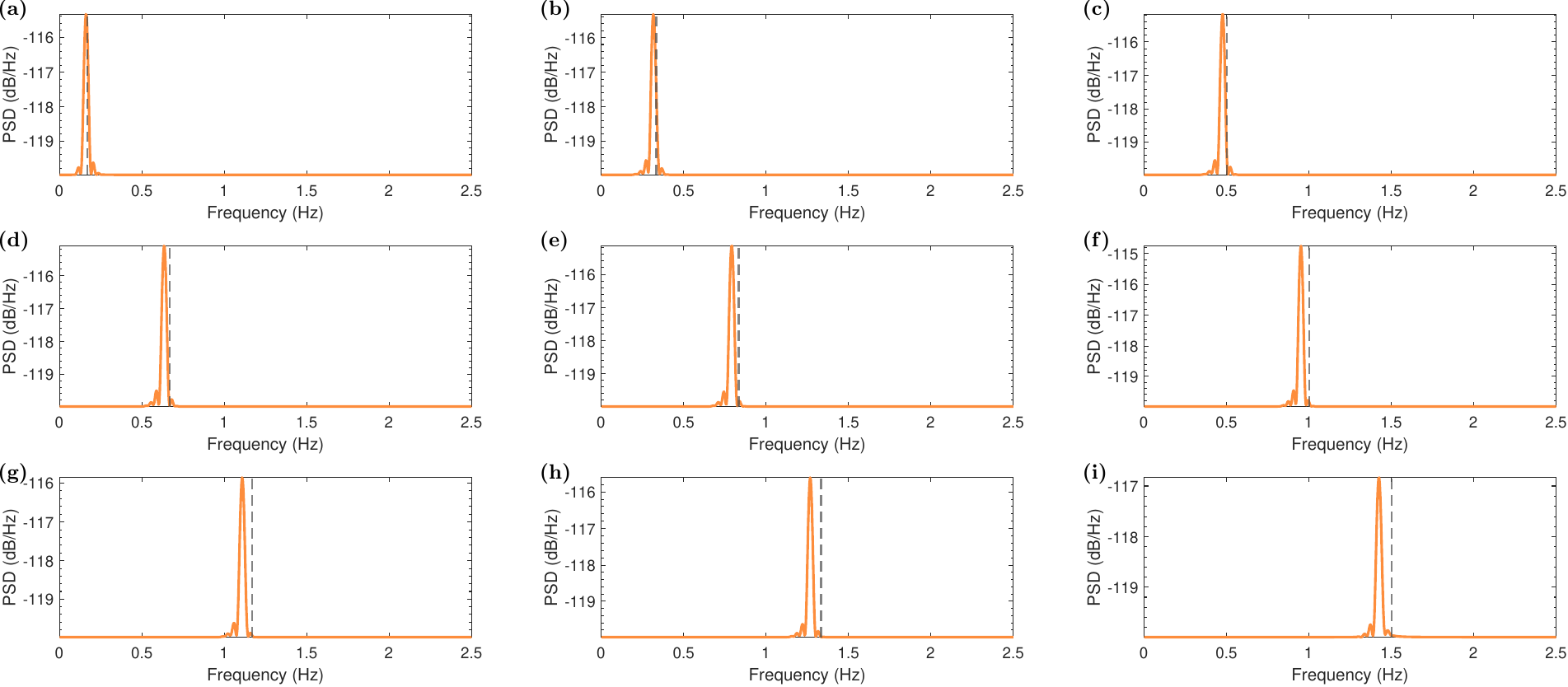}
		\caption{Power spectral density of Modes 1 through 9 extracted by TFMD. Subplots (a) through (i) correspond
			to Modes 1 through 9, respectively. Dashed vertical lines mark the nominal 1P frequencies.}
		\label{fig:comp1_freq}
	\end{figure}

	TFMD was applied to the preprocessed signal using the default parameter set, with the window shape parameter adjusted to $1.5$ to enhance frequency resolution. The first decomposition identified nine distinct modes. As illustrated in the time domain and spectral representations in \Cref{fig:comp1_time,fig:comp1_freq}, these modes (Modes 1--9) are mainly localized to the nine 30 s operating intervals. Their peak frequencies lie near the nominal 1P frequencies from 10 to 90 RPM, and their energies concentrate in the corresponding operating intervals. This supports interpreting the first decomposition as reconstructing modes from STFT support regions associated with the dominant 1P response under the stepped speed protocol.
	
\begin{figure}[htbp]
		\centering
		\includegraphics[width=0.95\textwidth]{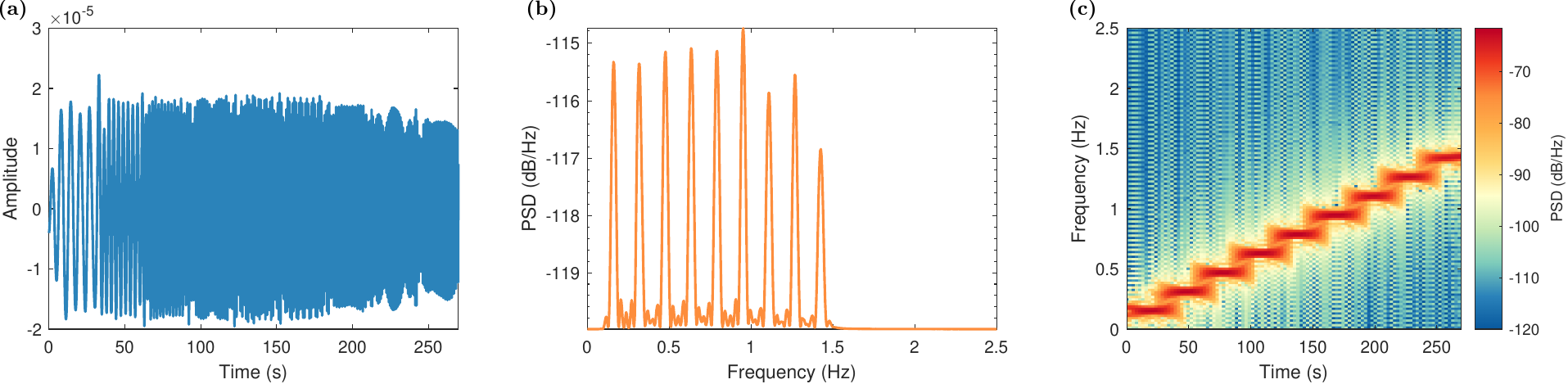}
		\caption{Analysis of the reconstruction after the first decomposition, $\hat{x}^{(1)}(t)$. (a) Time domain waveform. (b) PSD. (c) Spectrogram showing support regions near the nominal 1P frequencies.}
		\label{fig:recons1_analysis}
	\end{figure}

	The initial reconstruction, $\hat{x}^{(1)}(t)$, is obtained by summing these nine modes. As analyzed in \Cref{fig:recons1_analysis}, $\hat{x}^{(1)}(t)$ reconstructs modes from support regions near the nominal 1P frequencies. The first-stage residual was then decomposed to inspect weaker rotor-order structure.
	
\begin{figure}[htbp]
		\centering
		\includegraphics[width=0.8\textwidth]{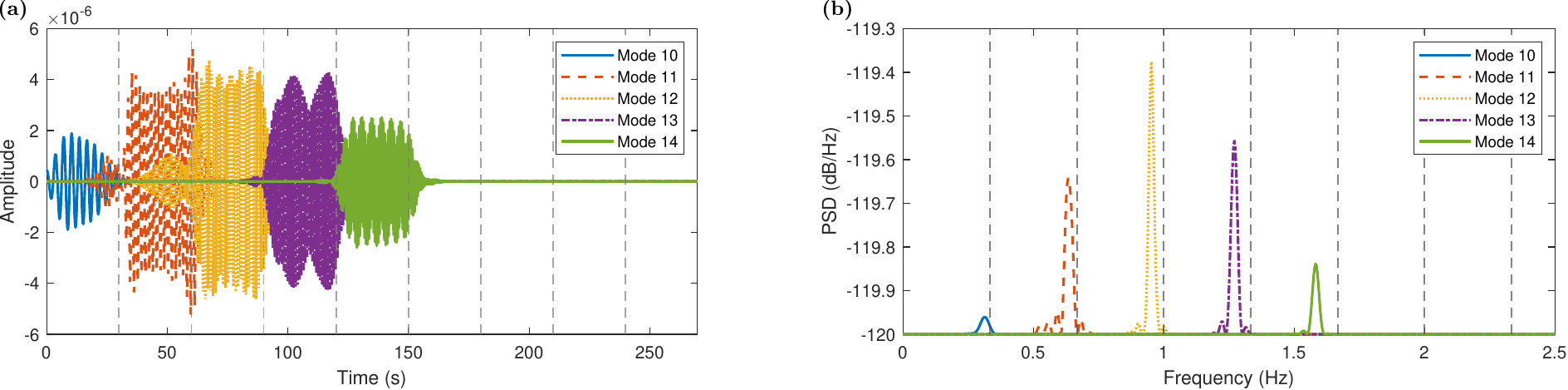}
		\caption{Analysis of the five residual modes (Modes 10--14) extracted from the residual signal after the first decomposition. (a) Overlaid time domain waveforms. (b) Overlaid PSD showing residual harmonic structure. Dashed
			vertical lines indicate 30 s operating intervals in subplot (a) and mark the nominal 2P frequencies in
			subplot (b).}
		\label{fig:comp2_analysis}
	\end{figure}

	TFMD was subsequently applied to the residual signal $e^{(1)}(t) = x(t) - \hat{x}^{(1)}(t)$. This residual decomposition identified five modes, reported as Modes 10 through 14 for continuity with the first decomposition. As shown in \Cref{fig:comp2_analysis}, their spectra show weaker harmonic peaks near the 2P frequencies at lower speeds, but the result should be interpreted as evidence that weaker harmonics can be detected in the residual rather than as complete 2P recovery or as a ground-truth decomposition of physical components.
	
\begin{figure}[htbp]
		\centering
		\includegraphics[width=0.95\textwidth]{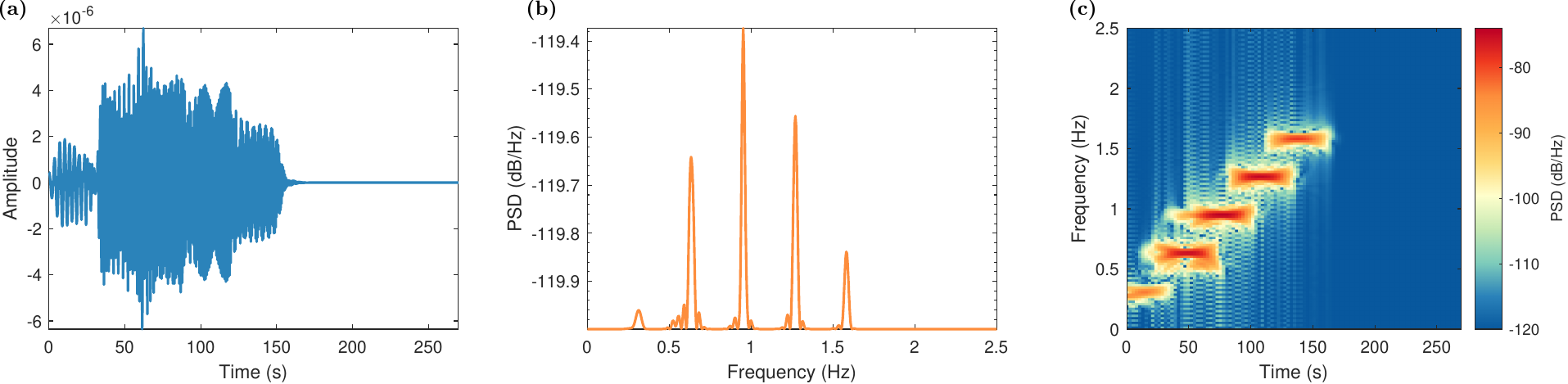}
		\caption{Analysis of the residual reconstruction $\hat{x}^{(2)}(t)$ from residual decomposition. (a) Time domain waveform. (b) PSD. (c) Spectrogram highlighting weaker residual harmonic structure.}
		\label{fig:recons2_analysis}
	\end{figure}

\begin{figure}[htbp]
		\centering
		\includegraphics[width=0.95\textwidth]{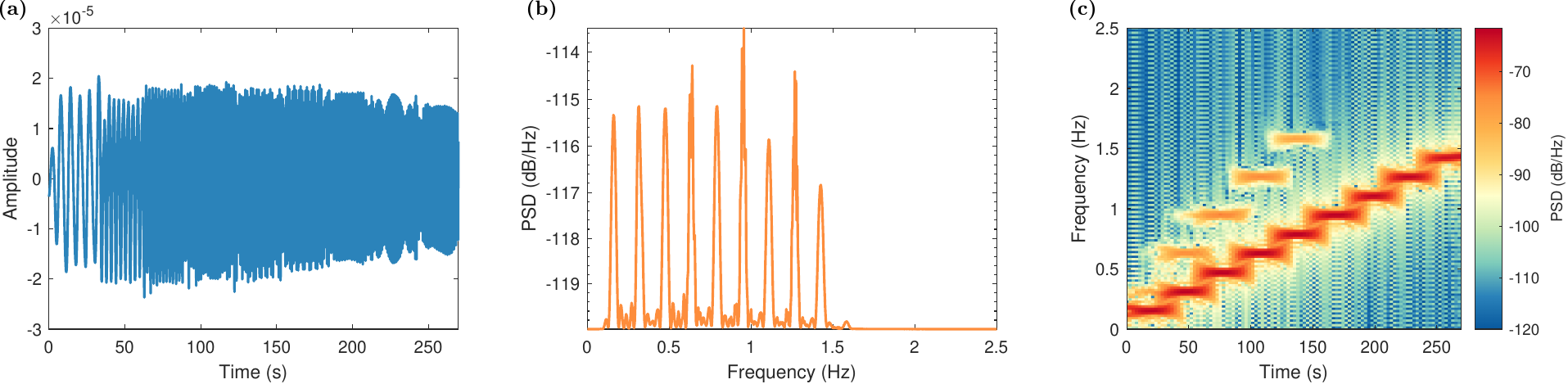}
		\caption{Analysis of the final reconstructed signal $\hat{x}_{\text{final}}(t)$. (a) Time domain waveform. (b) PSD. (c) Spectrogram showing support regions from the first decomposition near the nominal 1P frequencies together with harmonic structure extracted by residual decomposition.}
		\label{fig:recons_total_analysis}
	\end{figure}

	The signal reconstructed from the modes extracted from the residual signal, $\hat{x}^{(2)}(t)$, reduces the remaining residual structure (\Cref{fig:recons2_analysis}). The residual energy fraction decreases from 3.16\% after the first stage to 0.93\% after residual decomposition. The final reconstruction, $\hat{x}_{\text{final}}(t) = \hat{x}^{(1)}(t) + \hat{x}^{(2)}(t)$, is presented in \Cref{fig:recons_total_analysis}. The spectrogram shows the support regions from the first decomposition near the nominal 1P frequencies together with harmonic structure extracted by residual decomposition.
	
	\subsection{Comparative Analysis with Benchmark Methods}
	\label{subsec:application_comparison}
	
	To provide a comparative assessment, the preprocessed blade strain signal was decomposed using five established benchmark methods, namely VMD, EMD, ACMD, SET, and VGNMD. For methods that accept or use an output count setting, nine was supplied to align with the nine operating states. VMD used this value as the number of modes, ACMD and SET extracted nine components or ridges, and EMD used it as the maximum number of intrinsic mode functions. VGNMD used its internal mode detection.
	
	The benchmark results in \Cref{tab:wind_benchmark_summary} show different tradeoffs rather than uniform failure. The first TFMD decomposition provides a direct interpretation with respect to the operating intervals, because its modes contain spectral peaks near the nominal 1P frequencies and are concentrated within the corresponding intervals. VMD and SET produce outputs with peaks near most nominal 1P frequencies. However, VMD is less localized in the corresponding operating intervals, whereas SET has a high $\mathcal{E}_{\text{rel, total}}$ despite its good interval localization. VMD, EMD, and ACMD achieve lower $\mathcal{E}_{\text{rel, total}}$ values than the final two-stage TFMD reconstruction on this measured signal, but their outputs map less cleanly to the nine operating intervals. VGNMD returns 36 outputs, with peaks near several nominal 1P frequencies, which makes the decomposition harder to interpret.

	\begin{table}[htbp]
		\centering
		\caption{Benchmark summary for the controlled wind turbine blade strain signal.}
		\label{tab:wind_benchmark_summary}
		\small
		\begin{tabular}{lcc}
			\toprule
			Method & Number of outputs & $\mathcal{E}_{\text{rel, total}}$ \\
			\midrule
			TFMD (two-stage) & 14 (9+5) & 0.0966 \\
			VMD & 9 & 0.0945 \\
			EMD & 7 & 0.0467 \\
			ACMD & 9 & 0.0066 \\
			SET & 9 & 0.6098 \\
			VGNMD & 36 & 0.5520 \\
			\bottomrule
		\end{tabular}
	\end{table}
	
\begin{figure}[htbp]
		\centering
		\includegraphics[width=0.90\textwidth]{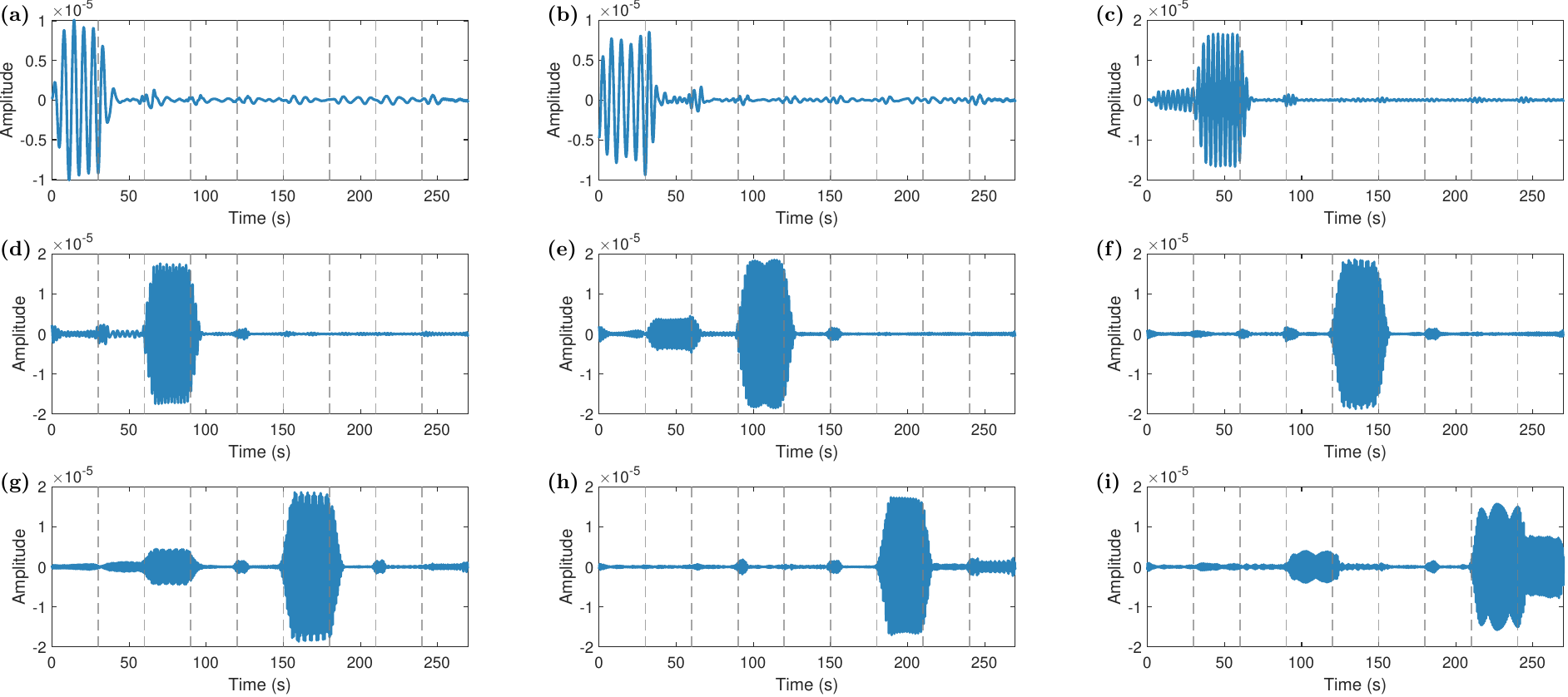}
		\caption{Time domain representations of Modes 1 through 9 extracted by VMD from the blade strain signal. Subplots (a) through (i) correspond to Modes 1 through 9, respectively. Dashed vertical lines indicate 30 s operating intervals.}
		\label{fig:vmd_time_domain}
	\end{figure}

\begin{figure}[htbp]
		\centering
		\includegraphics[width=0.90\textwidth]{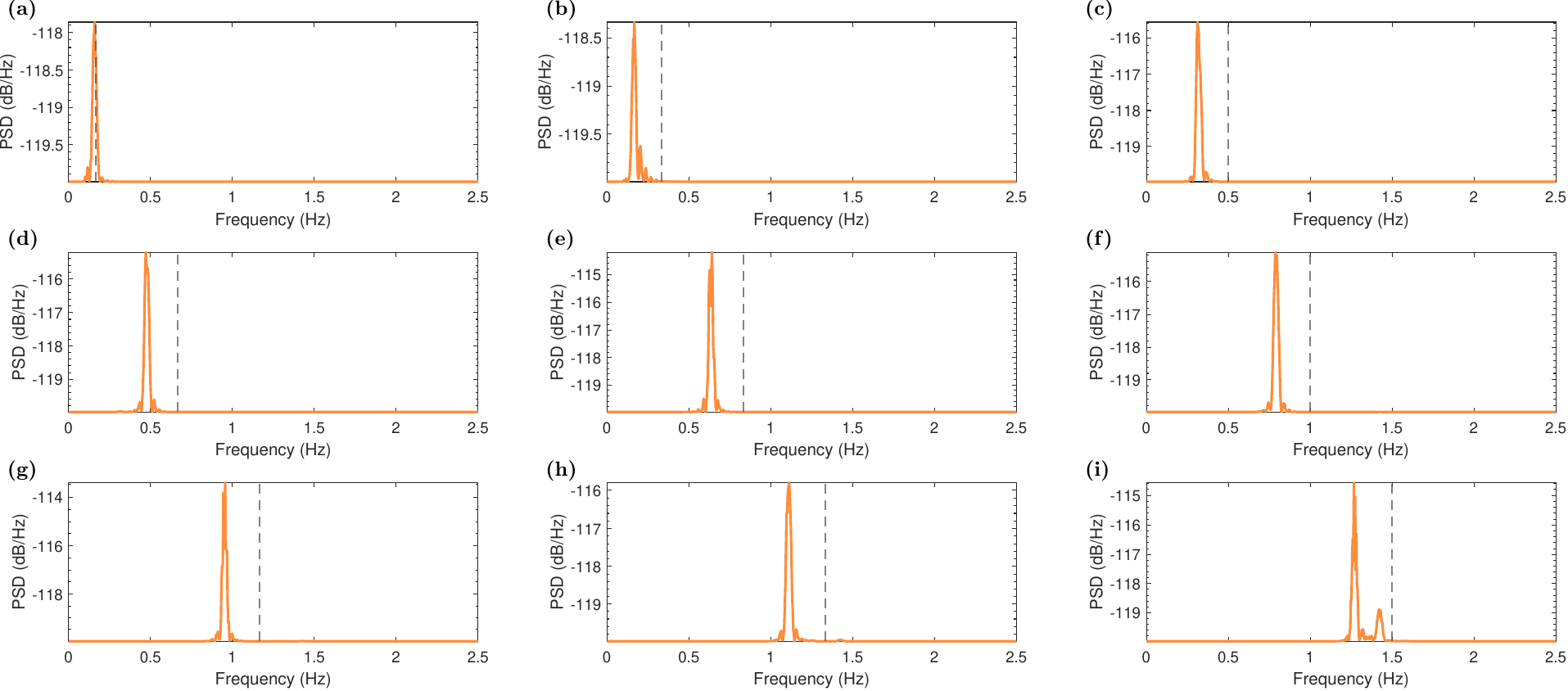}
		\caption{Power spectral density of Modes 1 through 9 extracted by VMD. Subplots (a) through (i) correspond to Modes 1 through 9, respectively. Dashed vertical lines mark the nominal 1P frequencies.}
		\label{fig:vmd_frequency_domain}
	\end{figure}

\begin{figure}[htbp]
		\centering
		\includegraphics[width=0.90\textwidth]{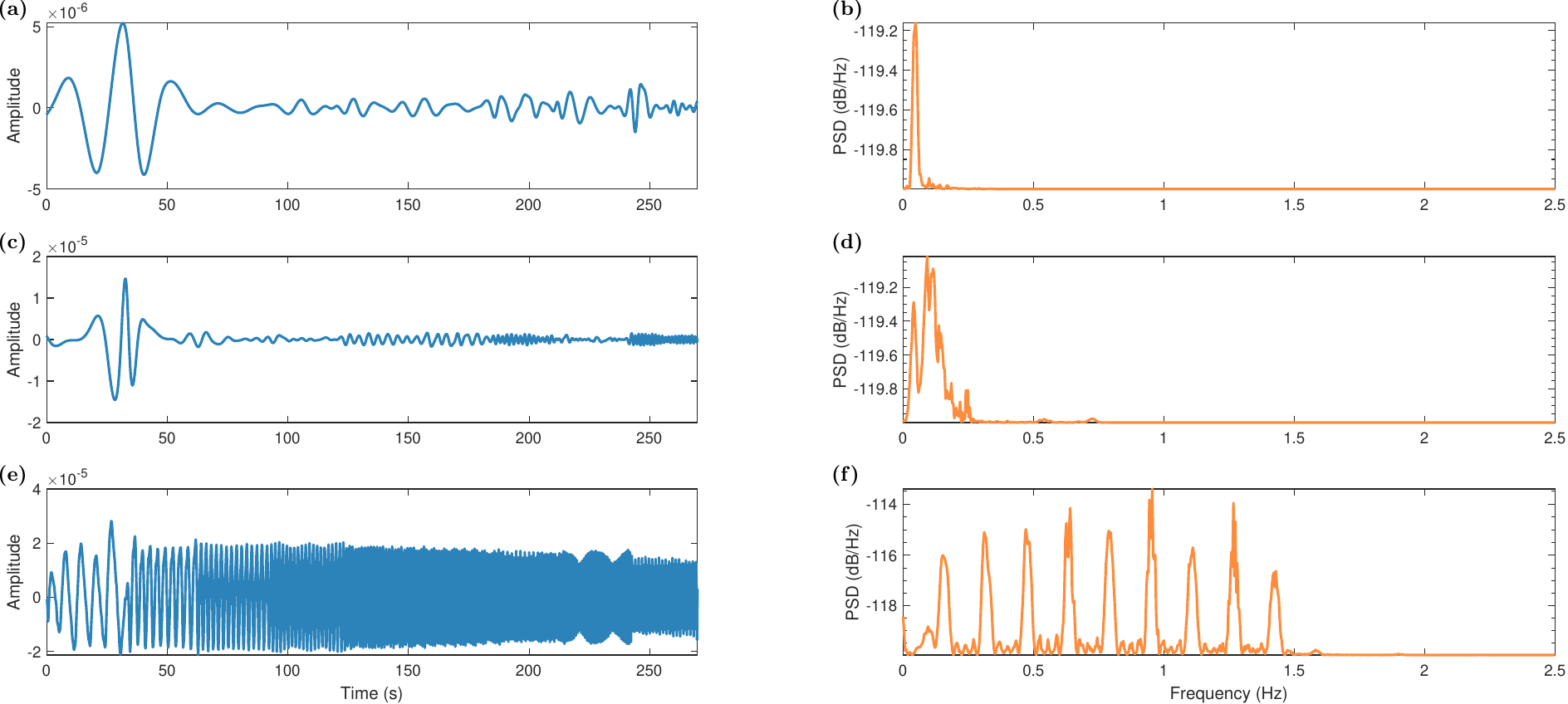}
		\caption{Decomposition results for the blade strain signal using EMD. Subplots (a), (c), and (e) show time domain waveforms, and (b), (d), and (f) show corresponding PSD for the three most energetic outputs.}
		\label{fig:emd_components}
	\end{figure}

\begin{figure}[htbp]
		\centering
		\includegraphics[width=0.90\textwidth]{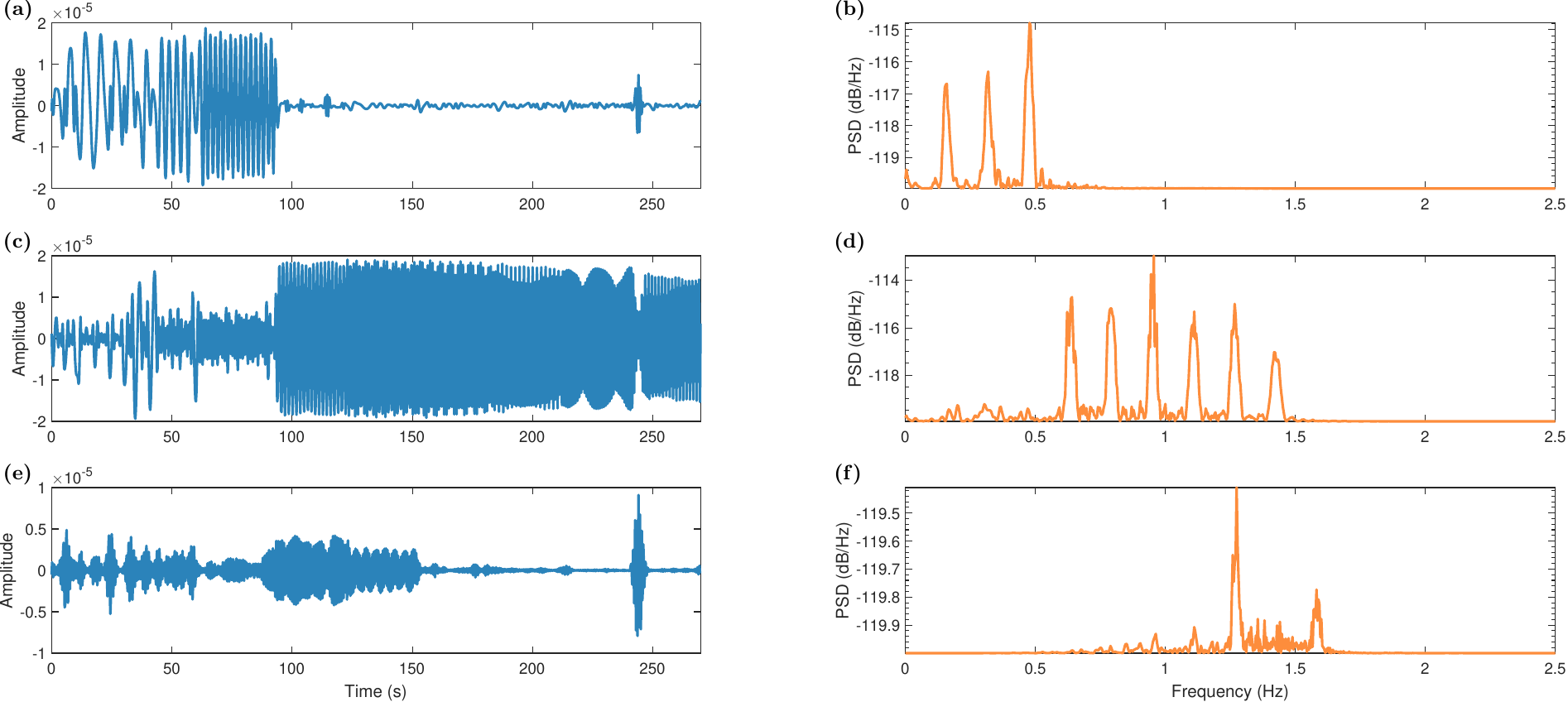}
		\caption{Decomposition results for the blade strain signal using ACMD. Subplots (a), (c), and (e) show time domain waveforms, and (b), (d), and (f) show corresponding PSD for the three most energetic outputs.}
		\label{fig:acmd_components}
	\end{figure}

\begin{figure}[htbp]
		\centering
		\includegraphics[width=0.90\textwidth]{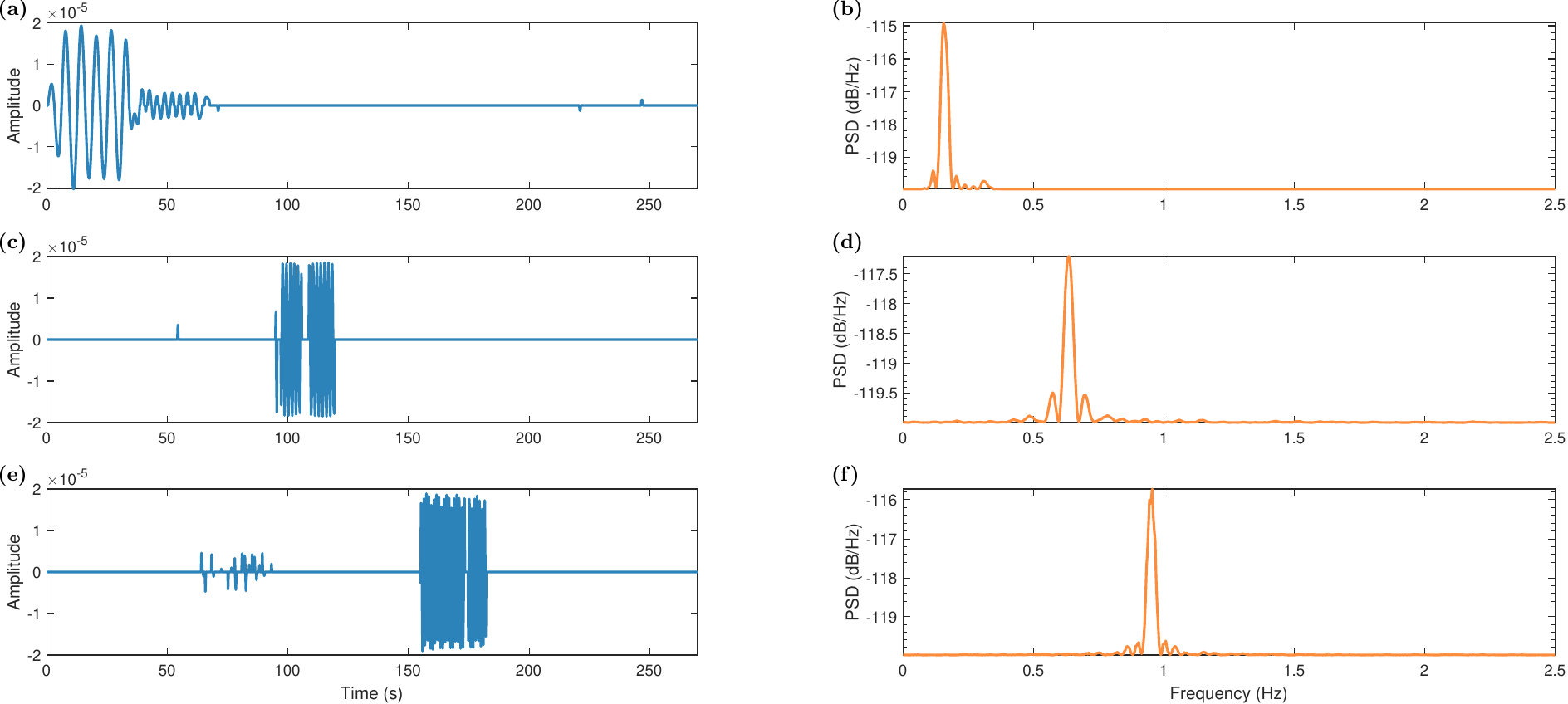}
		\caption{Decomposition results for the blade strain signal using SET. Subplots (a), (c), and (e) show time domain waveforms, and (b), (d), and (f) show corresponding PSD for the three most energetic outputs.}
		\label{fig:set_components}
	\end{figure}

\begin{figure}[htbp]
		\centering
		\includegraphics[width=0.90\textwidth]{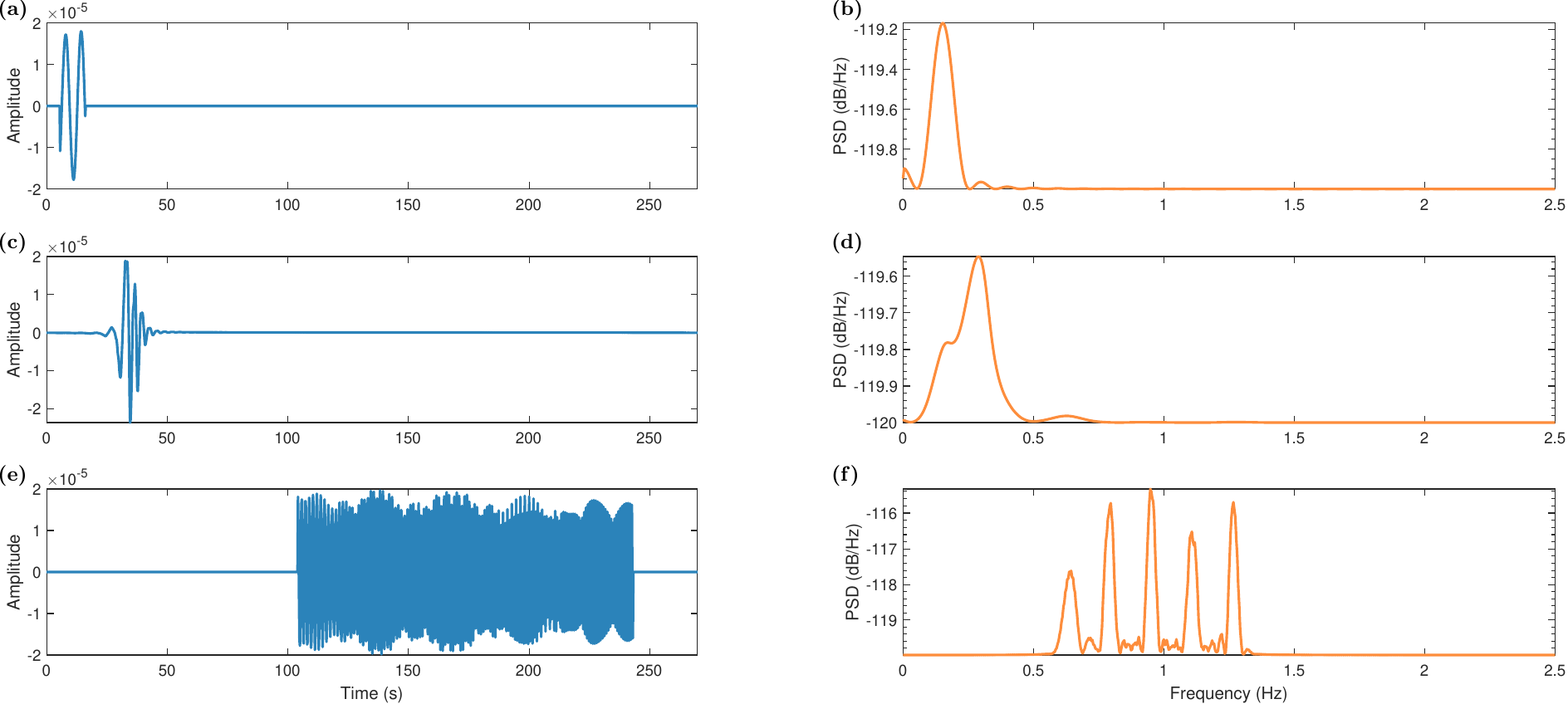}
		\caption{Decomposition results for the blade strain signal using VGNMD. Subplots (a), (c), and (e) show time domain waveforms, and (b), (d), and (f) show corresponding PSD for the three most energetic outputs.}
		\label{fig:vgnmd_components}
	\end{figure}

	The corresponding figures illustrate how these tradeoffs appear in time and frequency. \Cref{fig:vmd_time_domain,fig:vmd_frequency_domain} show that VMD outputs have visible temporal spread across several operating intervals, while \Cref{fig:emd_components,fig:acmd_components} show EMD and ACMD outputs that are harder to align with the operating state sequence. SET retains localized responses near many operating intervals but remains limited by its high $\mathcal{E}_{\text{rel, total}}$ (\Cref{fig:set_components}). VGNMD extracts fragmented outputs, which makes the decomposition harder to interpret (\Cref{fig:vgnmd_components}). For compactness, the three most energetic outputs are shown for EMD, ACMD, SET, and VGNMD. These comparisons support the narrower conclusion that TFMD is not universally best on every metric, but it gives a favorable tradeoff here through a compact set of modes, spectral peaks near the nominal 1P frequencies, and modes that follow the prescribed sequence of operating intervals.
	
	\section{Discussion}
	\label{sec:discussion}
	
	The numerical experiments support TFMD as a practical decomposition method when components occupy separated support regions in the sampled STFT plane. In this setting, the method infers the number of modes from connected support regions rather than requiring it as a direct input, and reconstructs modes without training data or a prescribed parametric signal model. Across the tested synthetic cases, TFMD identifies the number of modes stably, its error decreases with increasing SNR, and its median $\mathcal{E}_{\text{rel, avg}}$ is lower than that of the closest benchmark in each reported case, with smaller margins in Cases 2 and 4 and larger margins in Cases 3, 5, and 6.
	
	The wind turbine blade strain experiment shows how this segmentation view can aid interpretation under stepped speed operation. The first decomposition identifies nine support regions near the nominal 1P frequencies, and their energies are concentrated in the corresponding operating intervals. Residual decomposition further reduces the remaining energy and indicates weaker harmonic structure, including peaks near lower speed 2P frequencies. This result should be read as evidence of residual structure related to rotor order, not as complete 2P recovery or physical structural modal identification. The decision to inspect the residual therefore remains an engineering choice guided by expected rotor-order frequencies and the purpose of the analysis.
	
	In these evaluations, decomposition accuracy and interpretability must be considered together. VMD and SET produce outputs with peaks near most nominal 1P frequencies. SET shows comparatively good interval localization but a high reconstruction error. VMD, EMD, and ACMD can yield low reconstruction errors in some cases, but these three methods are less consistently concentrated within the prescribed operating intervals in the laboratory signal. VGNMD tends to produce fragmented results. The numerical timing comparison shows that TFMD is not always the fastest method, but its combination of moderate runtime, connected support regions, and direct reconstruction from masks makes the output easier to relate to operating states when the support regions are separated.
	
	The main limitation is also clear from this formulation. TFMD depends on component separability in the selected STFT plane, as stated in Assumptions \ref{assum:separability} and \ref{assum:energy}. If two components remain closer than the effective STFT frequency resolution over a sustained interval, the initial connected region selection may merge their supports into one region, and TFMD then reconstructs them as one mode. The present wind turbine experiment was conducted on a controlled laboratory rotor with known nominal speed steps, so it does not establish offshore field use. Offshore measurements would require validation under hydrodynamic loading, turbulent inflow, yaw and pitch control, drivetrain interactions, changing sensor conditions, and long records under variable operating conditions, with independent operating logs or order references.

	\section{Conclusion}
	\label{sec:conclusion}
	
	This study shows that estimating connected support regions in the STFT plane provides a practical route to multicomponent vibration analysis under variable speed operation. Each mode is reconstructed from one connected support region, avoiding prescribed parametric models or manually selected ridges.

	Three findings emerge from the numerical evaluations. First, TFMD achieves the lowest median $\mathcal{E}_{\text{rel, avg}}$ across the tested SNR levels in all six synthetic cases, while inferring the number of modes from connected support regions within each decomposition stage. Second, the method maintains correct identification of the number of modes across the tested noise conditions with the default dilation factor $\beta = 0.5$. Third, under the median-of-five timing protocol, TFMD is not uniformly fastest, but its average runtime remains below 0.06 s in all six numerical cases and is second shortest in Cases 2--6.

	The laboratory wind turbine experiment indicates that TFMD can reconstruct modes from support regions near the nominal 1P frequencies in a controlled stepped speed blade strain response and can reveal weaker residual harmonic structure through residual decomposition. These results support TFMD as a candidate tool for vibration analysis under variable speed operation, but they do not establish complete 2P recovery, structural mode identification, or offshore field use. Future work should evaluate the method on offshore wind turbine measurements with independent operating references and report order tracking, and detection of weaker harmonics.
	
	\section*{CRediT authorship contribution statement}

	\textbf{Wei Zhou}: Conceptualization, Methodology, Software, Formal analysis, Investigation, Validation, Visualization, Writing – original draft, Funding acquisition.
	\textbf{Wei-jian Li}: Methodology, Formal analysis, Validation, Writing – review \& editing, Funding acquisition.
	\textbf{Desen Zhu}: Investigation, Methodology, Data curation, Writing – review \& editing.
	\textbf{Hongbin Xu}: Supervision, Resources, Funding acquisition, Writing – review \& editing.
	\textbf{Wei-xin Ren}: Conceptualization, Supervision, Resources, Project administration, Funding acquisition, Writing – review \& editing.
		
	\section*{Data availability}
	
	The source code developed in this study is publicly available at \url{https://github.com/dopawei/TFMD}. The experimental data used in this study are not publicly available due to project confidentiality constraints.
	
	\section*{Acknowledgments}

	The authors are grateful for the financial support from Shenzhen Science and Technology Program through Grant No. JCYJ20220818100202006 and JCYJ20240813143211014, National Natural Science Foundation of China through Grant No. 12402483 and China Postdoctoral Science Foundation under Grant No. GZC20252166.

	\section*{Conflicts of Interest}

	The authors declare no conflict of interest.

\end{document}